\documentstyle[12pt,amssymb,amscd]{amsart}  

\newtheorem{thm}{Theorem}[subsection]

\newtheorem{prop}[thm]{Proposition}
\newtheorem{lemma}[thm]{Lemma}

\newtheorem{definition}[thm]{Definition}
\newtheorem{proposition}[thm]{Proposition}

\theoremstyle{remark}
\newtheorem{remark}[thm]{Remark}
\newtheorem{example}[thm]{Example}

\theoremstyle{definition}

\numberwithin{equation}{section}

\newcommand{\bbA}{{\Bbb A}}
\newcommand{\bbC}{{\Bbb C}}

\newcommand{\bbR}{{\Bbb R}}

\newcommand{\bbK}{{\Bbb K}}
\newcommand{\bbZ}{{\Bbb Z}}

\newcommand{\Cxxi}{{\Bbb C}[[\widehat{x}^{1}, \ldots , \widehat{x}^{n},
, \widehat{\xi}^{1}, \ldots , \widehat{\xi} ^{n}, \hbar]]}

\newcommand{\fx}{\widehat{x}}
\newcommand{\ftheta}{\widehat{\theta}}
\newcommand{\fxi}{\widehat{\xi}}
\newcommand{\tH}{\widetilde{H}}
\newcommand{\eq}{\sim}

\newcommand{\R}{\operatorname{\bold R}}

\newcommand{\tA}{{\widetilde{A}}}

\newcommand{\isomoto}{\overset{\sim}{\to}}

\newcommand{\fot}{\frac{1}{2}}
\newcommand{\foh}{\frac{1}{i\hbar}}
\newcommand{\fih}{\frac{i}{\hbar}}
\newcommand{\tn}{\widetilde{\nabla}}
\newcommand{\tnabla}{\widetilde{\nabla}}
\newcommand{\g}{{\frak{g}}}

\newcommand{\tg}{\widetilde{\g}}
\newcommand{\tG}{\widetilde{G}}

\newcommand{\jets}{\operatorname{jets}}
\newcommand{\LA}{\operatorname{\Lambda}}
\newcommand{\TLA}{\tilde{\operatorname{\Lambda}}}
\newcommand{\TLAN}{\tilde{\operatorname{\Lambda}}^N}
\newcommand{\SP}{\operatorname{Sp}}
\newcommand{\TSP}{\widetilde{\operatorname{Sp}}}
\newcommand{\TSPN}{\widetilde{\operatorname{Sp}}^N}
\newcommand{\TU}{\tilde{U}}
\newcommand{\VHL}{V ^H _L}
\newcommand{\TUN}{\tilde{U}^N}
\newcommand{\DX}{D^{\frac{1}{2}}_X}

\newcommand{\RDX}{{\cal R}D^{\frac{1}{2}}_X}
\newcommand{\RDUa}{{\cal R}D^{\frac{1}{2}}_{U^0 _{\alpha}}}

\newcommand{\Ci}{C^{\infty}}
\newcommand{\Cip}{C^{\infty, \operatorname{poly}}}
\newcommand{\TX}{T^*X}
\newcommand{\foho}{\frac{1}{i\hbar}\omega}
\newcommand{\Ub}{U_{\beta}}

\newcommand{\Uoab}{U^{0} _{\alpha (\beta)}}
\newcommand{\Uoa}{U^{0} _{\alpha}}

\newcommand{\Ubo}{U_{\gamma}}
\newcommand{\Ug}{U_{\gamma}}
\newcommand{\ahat}{\widehat{a}_{\beta, \hbar}}
\newcommand{\aohat}{\widehat{a}_{\gamma, \hbar}}
\newcommand{\Rkhat}{\widehat{R_{\beta, \gamma, k}  (a)}}
\newcommand{\Rk}{R_{\beta, \gamma, k}}
\newcommand{\Tkhat}{\widehat{T_{\alpha, \alpha _{1}, k}  (f)}}
\newcommand{\Tk}{T_{\alpha, \alpha _{1}, k}}
\newcommand{\Pkhat}{\widehat{P_{\beta, k}  (f,a)}}
\newcommand{\Pk}{P_{\beta, k}}
\newcommand{\Dkhat}{\widehat{D_{\alpha, k}  (f_1, f_2)}}
\newcommand{\Dk}{D_{\alpha, k}}
\newcommand{\Ohi}{O(\hbar ^{\infty})}

\newcommand{\fhat}{\widehat{f}_{\alpha, \hbar}}
\newcommand{\fohat}{\widehat{f}_{\alpha _1, \hbar}}
\newcommand{\pijd}{\pi ^*{\operatorname{jets}}{\RDX}}
\thanks
{The second author is supported
in part by NSF grant 0830-300-L830.}
\makeindex
\begin{document}

\bigskip

{\centerline {\bf {Remarks on modules over deformation quantization
algebras}}}

\bigskip

{\centerline {Ryszard Nest and Boris Tsygan}}

\bigskip

\author{Ryszard Nest}
\address{Mathematical Institute, Copenhagen University, Universitatsparken,
2100 Copenhagen, Denmark} \email{rnest@@math.ku.dk}
\author{Boris Tsygan}
\address{Department of Mathematics, Northwestern University,
Evanston, IL, 60208, USA} \email{tsygan@@math.nwu.edu}

\bigskip

\centerline{To Boris Feigin on his fiftieth birthday}

\bigskip

\section{Introduction}

The aim of this paper is to provide a link between deformation
quantization theory of \cite{BFFLS} and \cite{Fe} and Lagrangian analysis. By the latter we mean Maslov's theory of canonical operator \cite{M}, \cite{MSS}, \cite{NSS}, \cite{L} and 
H\"{o}rmander's theory of distributions given by oscillatory
integrals \cite{GS}, \cite{H}. Though it was always clear that
such links exist (cf., for example, \cite{Kar}), and though the creators of deformation quantization were probably partially motivated by Lagrangian analysis, we are not aware of any works that relate the two subjects
explicitly.

Here are three reasons why, in our view, such a link may be
desired. First of all, there is a pedagogical reason: it is
natural to look for a more unified approach to the two important
subjects that are clearly related. Secondly, there is a motivation
from index theory. Namely, one can try to extend the Atiyah-Singer
index theorem from pseudo-differential operators to a more general
class of so called Guillemin-Sternberg operators which are Fourier
integral operators of special kind \cite{GS1}, \cite{GU}. (This is the authors' joint
project with A.~Gorokhovsky). To prove such a theorem, one would
try to reduce it to a general index theorem for deformation
quantization like in \cite{BNT}, \cite{NT}. For that, one needs
to answer questions which are studied in the present paper: how to
relate Lagrangian analysis to deformation quantization and, more
precisely, how to pinpoint the resulting algebra in terms of the
general classification of deformation quantizations of a
symplectic manifold given by the construction of Fedosov.

Finally, our third motivation is related to mirror symmetry. It is
a general feeling among the experts that the Fukaya category of a
symplectic manifold is somehow, in a very nontrivial way, related
or analogous to the derived category of modules over a deformation
quantization of this manifold (cf., for example, \cite{BS} or
\cite{KS}; the idea that deformation quantization should be
related to Lagrangian intersection theory was communicated to the
second author by Boris Feigin in the mid 80s). We think that our
constructions may suggest new structures on modules over deformed
algebras, which would lead to modified versions of the derived
category of modules which might be related to the Fukaya category
somewhat more closely.

More precisely, when our symplectic manifold has the first Chern
class equal to zero, then its deformation quantization leads to an
additional structure, namely to a groupoid $\tG$ which, roughly
speaking, consists of expressions $\exp (\fih H)$ where $H$ is a
function (cf. \ref{ss:groupoid tG}). Our feeling is that the kinds
of modules which appear in deformation theory from Lagrangian analysis are something like objects of a new derived category of
complexes of locally free modules; in that new category,
localizing with respect to quasi-isomorphisms is modified so that
elements of $\tG$ are included into the set of quasi-isomorphisms
by which we allow to localize. Such a construction could be a
somewhat better approximation to the Fukaya category because it is
not local, i.e. not purely sheaf-theoretical. Also, those modules
would be in a closer relation to the Maslov phenomena which are
cental to Lagrangian intersections \cite{Se} but cannot be seen by
the ordinary homological algebra of modules over deformation
quantizations.

Let us describe the contents of the paper in more detail. After
some preliminaries on Lagrangian subspaces, Lagrangian
submanifolds, the Maslov index, and an algebraic version of the Weil representation, we review the Fedosov construction
and classification of deformation quantizations. Then we remind
how to construct a deformation quantization of $\TX$ starting from
differential operators on $X$. We then compute this deformation
quantization in terms of Fedosov's classification \cite{Fe}. Our
version of this construction essentially follows \cite{BNT}, but
we design a modified Fedosov construction which streamlines the
exposition. More precisely, the Fedosov construction provides a
deformation quantization starting from a multiplication preserving
connection on the Weyl bundle of a symplectic manifold. We show
that differential operators on half-densities lead to a product
which is defined directly on the bundle of jets. This product is
preserved by the canonical connection on the jet bundle. This is
the canonical bundle of algebras ${\cal W}$ which is isomorphic to the Weyl bundle $W$ of Fedosov. This isomorphism is canonical up to a canonical connection-preserving inner automorphism. (This means, here and below, that the isomorphism canonically depends on a choice of an auxiliary datum; isomorphisms corresponding on two different choices differ by a conjugation by an element which is canonically constructed from the pair of data. Next we review Lagrangian analysis, in particular H\"{o}rmander's construction of distributions whose
wave front is a given Lagrangian submanifold $L$. We show that,
after an extension of the ring of scalars, the asymptotics of this
construction leads to a module over the deformation quantization
of $\TX$ discussed above. Our exposition here is close to Maslov's method of canonical operator, cf. \cite{NSS}.

We would like to express this module in more familiar
deformation-theoretic terms. There are two equivalent ways of
doing that. First, one can express it in terms of the Fedosov
construction of deformations of symplectic manifolds. Second, we
can apply Darboux-Weinstein theorem and identify an open
neighborhood of $L$ in $T^*X$ with an open neighborhood of $L$ in
$T^*L$. By the classification theorem for deformations of a symplectic manifold, the deformed algebra
on $T^*X$, restricted to the neighborhood, becomes isomorphic to
the standard deformation on $T^*L$. Moreover, as we show in  \ref{prop:can up to in}, this isomorphism is canonical up to a canonical inner automorphism. We prove that, after identifying the two deformed algebras using this isomorphism, the Lagrangian module corresponding to $L$ in $T^*X$ becomes
isomorphic to the similar module corresponding to $L$ in
$T^*L$, tensored by the flat bundle given by a certain \v{C}ech
one-cocycle (Theorem \ref{thm:main}). This cocycle involves the
Maslov class of $L$ and the cohomology class of $\alpha |L$ where
$\alpha$ is the standard one-form on $T^*X$ such that $d\alpha =
\omega$.

Our key observation is that modules of the type we consider are
still perfectly well described by Gelfand's formal differential
geometry. Namely, one can construct the bundle of jets of sections
of such a module, which is a bundle of modules over the algebra of
jets of functions. This is, in a sense, a second
microlocalization: after having localized the distributions to a
Lagrangian submanifold, we now further localize them to any point
of this submanifold.

Once the jet formalism for our modules is established, one can compare them to each other. More precisely, using local phase functions of $L$, we see that the Lagrangian jet bundle is isomorphic to the vector bundle induced by the algebraic Weil representation of the universal cover of the symplectic group. The isomorphism is given, essentially, by the Maslov canonical operator at the jet level. This bundle, in turn, is easy to compare to the Lagrangian jet bundle of $L$ in $T^*L$.

We believe that most of the contents of the paper are well known
to experts in some form. The second author is greatly indebted to
Boris Feigin for introducing him to the topic (and to deformation
quantization in general, as well as to many other things). He is
also grateful to Alexander Karabegov for sharing a key idea how
to establish a direct connection between the stationary phase
method and deformation quantization. We are grateful to B. Sternin for a masterful exposition of the theory of canonical operator. We are grateful to D. Arinkin, A. Beilinson, R. Bezrukavnikov, P. Bressler, V. Drinfeld, K. Fukaya, D. Gaitsgori, A. Gorokhovsky, M. Kashiwara, D. Kazhdan, Yu. Manin, and D. Tamarkin for helpful remarks and discussions.
\section{Preliminaries from symplectic linear algebra} \label{section:linear
algebra}
\subsection{}
Let $\bbR ^{2n}$ be the standard symplectic vector space with the
symplectic form $d\xi \wedge  dx = \sum d\xi ^k \wedge dx ^k$. Let $\LA$ be
the set of Lagrangian subspaces of $\bbR ^{2n}$. One has $U(n)/O(n)
\isomoto \LA$. It is well known that the map
$$ u \in U(n) \mapsto \det (u)  ^{2} \in S ^{1}$$
induces an isomorphism $\pi _1 (\LA) \isomoto \pi _1(S ^{1}) = \bbZ$.

For $N \in \bbZ$, $N > 0$, put

$$ \TUN (n) = \{(u, \zeta) | u \in U(n), \zeta \in \bbC, \det(u) ^{2} =
\zeta ^N\}
$$

This is a central extension of $U$ by $\bbZ / N$. The space $\TLAN
= \TUN / O$ is a cover of $\LA$ with the deck transformation group
$\bbZ/N$. Put also

$$ \TU (n) = \{(u, x) | u \in U(n), x \in \bbR, \det(u)^{2} = \exp(2\pi
ix)\}
$$
The space $\TLA  = \TU / O$ is the universal cover of $\LA$.

Let us describe a \v{C}ech one-cocycle determining the covering $\TLA \to
\LA$. For $I \subset \{1,\ldots , n\}$, let $ { x}_1 = (x^k | k \in I)$, $ {
x}_2 = (x^k | k \in {\overline {I}})$, $ { \xi }_1 = (\xi^k | k \in I)$, $ {
\xi} _2 = (\xi ^k | k \in {\overline {I}})$ where ${\overline{I}} $ is the
complement of $I$. Let $L_I$ be the Lagrangian subspace $\{ { x}_2 = 0,\, {
\xi}_1 = 0\}$. Let $U _I$ be the open subset of those $L$ for which the
projection onto $L _I$ along $L _{\overline {I}}$ is an isomorphism. In
other words, $U_I$ consists of those $L$ which are defined by equations
\begin{equation} \label{eq: parametrization of l_I 1}
\xi _1 = Ax_1 + B \xi _2
\end{equation}
\begin{equation} \label{eq: parametrization of l_I 2}
x _2 = -B^t x_1 - C \xi _2
\end{equation}
where $A$, $C$ are self-adjoint matrices and $t$ means transposition.

Now let us describe intersections $U_I \cap U_{J}$. Let $I = I_3 \cup I_4$,
$J = I_2 \cup I_4$, where $I_p$, $ p=1, \ldots, 4$, are disjoint and cover $
\{1,\ldots , n\}$. Put $x_p = (x^k | k \in I_p )$ and $\xi _p = (\xi ^k | k
\in I_p )$. A Lagrangian subspace $L$ is in $U_I \cap U_{J}$ if and only if
it can be described by equations
\begin{eqnarray}
\xi _1 &  = & \ldots \nonumber\\
\xi _2 & = & \ldots + Ax_2 + B\xi _3 + \ldots \nonumber\\
x_3 & = & \ldots  -B^t x_2 - C \xi _3 + \ldots \nonumber\\
x _4 & = & \ldots  \nonumber
\end{eqnarray}

where the matrix
$\left[ \begin{array}{cc} A& B\\ ^tB& C
\end{array} \right]$ is nondegenerate.
\begin{thm} \label{thm: Maslov cocycle, linear case}
( \cite{H}, \cite{GS}). The formula
$$c_{IJ} = \frac{1}{2}\operatorname{signature}\left[ \begin{array}{cc} A&
B\\ B^t& C
\end{array} \right]$$
defines a $\frac{1}{2} \bbZ$-valued \v{C}ech 1-cocycle for the cover
$\{U_I\}$. This cocycle is cohomologous to a $\bbZ$-valued 1-cocycle whose
cohomology class is the class of the universal cover of $\LA$.
\end{thm}
Here by the signature of a self-adjoint nondegenerate matrix we mean the
signature of the corresponding quadratic form.

There is a central extension $\TSPN$ of $\SP$ by $\bbZ /N\bbZ$ of which
$\TUN$ is a subgroup. It is defined by
$$\TSPN = \{(g,\gamma)\}$$
where $g \in \SP$ and $\gamma$ is a homotopy class of a path in
$\LA$ connecting $L_0$ and $g (L_0)$. Similarly, one
constructs a central extension $\TSP$ of $\SP$ by $\bbZ$.
\subsection{The Weil representation} \label{sss:weil}Here we recall in a slightly more algebraic form the standard construction of the Weil representation (cf., for example,\cite{L} and \cite{GS}). Let $V^{\operatorname{Weil}}$ be the vector space 
$$V^{\operatorname{Weil}} = \bigoplus _{T} \exp(\frac{iT\fx ^2}{2\hbar}){\bbC}[[\fx^1, \ldots, \fx^n, \hbar]],$$
i.e. a free ${\bbC}[[\fx^1, \ldots, \fx^n, \hbar]]$-module with the basis indexed by all complex symmetric $n\times n$ matrices $T$ such that $\operatorname{Im} T$ is positive definite.The group $\TSP ^4$ acts on $V^{\operatorname{Weil}}$ as follows. Let 
$$F(\fx) =  \exp(\frac{iT\fx ^2}{2\hbar})f(\fx^1, \ldots, \fx^n, \hbar)$$
where $f$ is a formal power series. Then
$$\left[ \begin{array}{cc} 1&
A\\ 0& 1
\end{array} \right]: F(\fx) \mapsto \exp(\frac{iA\fx ^2}{2\hbar})F(\fx)$$
for a real symmetric $n\times n$ matrix $A$;
$$\left[ \begin{array}{cc} B&
0\\ 0&{ B^{-1}}^t
\end{array} \right]: F(\fx) \mapsto |{\operatorname{det}}B|^{-\fot}F(B^{-1}\fx)$$
for $B \in {\operatorname{GL}}(n, \bbR)$;
$$\left[ \begin{array}{cc} 0&
1\\ -1& 0
\end{array} \right]: F(\fx) \mapsto {\operatorname{Fourier}}F(\fx);$$
the generator of the center of $\TSP ^4$ acts by multiplication by the imaginary unit $i$.
Here $ {\operatorname{Fourier}}$ stands for the Fourier transform at the level of power series as explained in \cite{K}; cf. \cite{L}, \cite{GS} for a related definition of the Fourier transform of asymptotics. Namely,
$${\operatorname{Fourier}}  \exp(\frac{iT\fx ^2}{2\hbar})f(\fx, \hbar)=f(\fxi, \hbar) {\operatorname{Fourier}}  \exp(\frac{iT\fx ^2}{2\hbar})$$
where $\fxi _k = i\hbar {\frac{\partial}{\partial \fx ^k}}$
and 
$$ {\operatorname{Fourier}}  \exp(\frac{iT\fx ^2}{2\hbar}) = (\operatorname{det}(-iT ^{-1}))^{-\fot}\exp(-\frac{iT^{-1}\fx ^2}{2\hbar})$$
Since the imaginary part of $T$ is positive definite, the square root is well defined.

There is also a degenerate version of the Weil representation:
$$V^{\operatorname{Weil}}_0 = \bigoplus _{T} \exp(\frac{iT\fx ^2}{2\hbar}){\bbC}[[\fx^1, \ldots, \fx^n, \hbar]],$$
where the sum is now taken over all real symmetric $n \times n$ matrices $T$. On this space, the representation is only partially defined. Namely, on the subspace $V_{T=0}$ one can define operators corresponding to elements of the open dense subset of $\TSP ^4$ whose projection to $\SP$ consists of matrices 
$$\left[ \begin{array}{cc} 1&
A\\ 0& 1
\end{array} \right]
\left[ \begin{array}{cc} 0&
B\\ {B^{-1}}^t& 0
\end{array} \right] 
\left[ \begin{array}{cc} 1&
C\\ 0& 1
\end{array} \right]$$
where $A$ is nondegenerate. On the other hand, the operators corresponding to elements whose projection is equal to 
$$\left[ \begin{array}{cc} B&
0\\ 0&{ B^{-1}}^t
\end{array} \right]$$
are defined everywhere.
Those elements whose action is defined on the subspace corresponding to an arbitrary $T$ belong to the above set conjugated by 
$\left[ \begin{array}{cc} 1&
T\\ 0& 1
\end{array} \right]$. Now one has to be more careful about the square root. We define for a real symmetric nondegenerate matrix $T$
$$(\operatorname{det}(iT))^{\fot} = \prod _{k=1}^n {\sqrt{i\lambda _k}}$$
where $\lambda _k$ are the eigenvalues and we take the branch of the square root$${\sqrt{re^{i\varphi}} } = {\sqrt r} e^{i\varphi /2}$$
for $-\pi <\varphi <\pi$. 

If the action of elements $g$, $h$, and $gh$ are defined on a vector, then the latter is equal to the composition of the former two. This can be seen by passing to the limit $T \to 0$ from $V^{\operatorname{Weil}}$ to $V^{\operatorname{Weil}} _0$.
\section{Preliminaries from symplectic geometry}
\subsection{} Let $(M, \omega)$ be a symplectic manifold. An
$\TSPN$-structure on $M$ is by definition a reduction of the
structure group of the tangent bundle $T_M$ to $\TSPN$. The group
$H ^{1}(M, \bbZ /N)$ acts transitively and freely on the set of
isomorphism classes of such reductions. An $\TSPN$-structure on
$M$ exists if and only if the image of $2c_1(T_M)$ in $H ^{2}(M,
\bbZ /N)$ is equal to zero. Here $c_1(T_M)$ is the first Chern
class of the tangent bundle viewed as a complex vector bundle
(after choosing an almost complex structure compatible with the
symplectic form).

An equivalent definition of a $\TSPN$-structure is as follows. Let
$\LA _M$ be the bundle whose fiber over a point $x$ is the
Grassmannian of Lagrangian subspaces of $T_x M$. To give an
$\TSPN$-structure on $M$ is the same as to give a bundle $\TLAN
_M$ with fiber $\TLA$, together with a morphism of bundles $\TLAN
_M \to \LA _M$ which is, at the level of the fibers, the morphism
$\TLAN  \to \LA $ (cf. \cite{Se}).

If a distribution of Lagrangian subspaces is given on $M$, then one can
define an associated $\TSPN$-structure as follows: the distribution provides
a base point in every $\LA _x$, and one uses this base point to define
$\TLAN  _x$. In the language of transition functions, observe that a distribution by Lagrangian subsets defines a reduction of the structure group from $\SP$ to the subgroup stabilizing a fixed Lagrangian submanifold $\bbR ^n$ of $\bbR ^{2n}$. But this subgroup admits a canonical lifting to a subgroup of $\TSP $, $g \mapsto (g,\:\gamma)$ where $\gamma $ is the constant path.

Given a Lagrangian submanifold $L$ of $M$, and given an $\TSPN$-structure on
$M$, one defines {\em the Maslov class} of $L$ as follows. On a tubular
neighborhood of $L$, there are two $\TSPN$-structures. One is the
restriction of the structure on $M$, the other comes from a Lagrangian
distribution which is transverse to $L$ (its isomorphism class does not
depend on a choice of such distribution). The two differ by an element of
$H^ {1} (L, \bbZ /N)$ which we call the Maslov class. Similarly, for an
$\TSP$-structure on $M$ and for a Lagrangian submanifold $L$, we define the
Maslov class in $H^ {1} (L, \bbZ )$.

Let us now describe the Maslov class of a Lagrangian submanifold
of $T^*X$ (where the $\TSP$-structure comes from the distribution
consisting of tangent spaces to fibers of the projection $T^* X \to X$) by a \v{C}ech one-cocycle of $L$. Let $\pi : T^* X \to
X$ be the projection. Consider an open cover $X = \cup U^0
_{\alpha}$ and a refinement $T^* X = \cup U_{\beta}$ of the cover
by $\pi ^{-1} ( U^0 _{\alpha})$. Every $U_\beta$ is contained in
some $\pi ^{-1} ( U^0 _{\alpha (\beta)})$. Using notation of
section \ref{section:linear algebra}, subdivide the coordinates on
each $ U^0 _{\alpha(\beta)}$ into two groups,
\begin{equation} \label{eq:local subdivision}
x = (x_1, x_2)
\end{equation}
in such a way that $L \cap U_{\beta}$ is given by equations
\begin{eqnarray}\label{local parametrization1}
\xi_1 & = & F_{x_1}(x_1, \xi _2) \\
x_2 & = & -F_{\xi _2} (x_1, \xi _2)
\end{eqnarray}
For an intersection $U_{\beta} \cap U_ {\gamma}$, our data can differ in two
ways.

1) They may differ by a choice of coordinates on $X$.

2) They may differ by a choice of subdivision \ref{eq:local subdivision} for
the same coordinate system.

Now let us start to define a one-cocycle representing the Maslov class. In
case 1), put
\begin{equation}\label{eq:Maslov cocycle 1}
c_{\beta \gamma} = 0
\end{equation}
In case 2), using notation of section \ref{section:linear
algebra}, let $I = I_3 \cup I_4$ for $U_\beta$ and $I = I_2 \cup
I_4$ for $U_\gamma$. On the intersection, $L$ can be described by
equations
\begin{eqnarray}\label{local parametrization2}
\xi_1 & = & F_{x_1}(x_1, x_2, \xi _3, \xi _4) \\
\xi_2 & = & F_{x _2} (x_1, x_2, \xi _3, \xi _4)\\
x_3 & = & -F_{\xi _3}(x_1, x_2, \xi _3, \xi _4)\\
x_4 & = & -F_{\xi _4} (x_1, x_2, \xi _3, \xi _4)
\end{eqnarray}
where the Hessian matrix $\operatorname{Hess}_{{\xi_2}, {x_3}}(F)$ is
nondegenerate. Put
\begin{equation}\label{eq:Maslov cocycle 2}
c_{\beta \gamma} = \frac{1}{2}
\operatorname{signature}\operatorname{Hess}_{{\xi_2}, {x_3}}(F)
\end{equation}
\begin{thm} \label{thm:Maslov cocycle 2}
(\cite{H}, \cite{GS}) The cochain $c$ is a $\frac{1}{2} \bbZ$ - valued
\v{C}ech one-cocycle of $L$ which is cohomologous to a $\bbZ$-valued cocycle
representing the Maslov class.
\end{thm}
The proof will be contained in the proof of Theorem \ref{thm:main}.
\section{Preliminaries from deformation quantization}
\subsection{}
Let $(M, \omega)$ be a symplectic manifold. A {\em deformation quantization}
of $M$ (cf. \cite{BFFLS}) is a formal power series
\begin{equation} \label{eq:deformation}
f*g = fg + \sum_{k=1}^{\infty} (i\hbar)^ k D_k (f,g)
\end{equation}
where $D_k : C^{\infty}(M) \otimes C^{\infty}(M) \to C^{\infty}(M)$ are
bilinear bidifferential expressions, $*$ is associative, $f*1 = 1*f = f$,
and
\begin{equation} \label{eq:Poisson bracket}
\{f,g\} = D_1 (f,g) - D_1 (g,f)
\end{equation}
is the Poisson bracket defined by the symplectic structure.

\begin{definition} \label{dfn:isomorphism}
An {\em isomorphism} between $*$ and $*'$ is a formal series
\begin{equation} \label{eq:isomorphism of deformations}
T(f) = f + \sum_{k=1}^{\infty} (i\hbar)^ k T_k (f)
\end{equation}
where $T(f*g) = T(f) *' T(g)$ and $T_k$ are differential operators on
$C^{\infty} (M)$.
\end{definition}
\begin{example} \label{example:Weyl}
For $M = \bbR ^{2n}$ and $\omega = d\xi \wedge dx$, put
\begin{equation} \label{eq:Moyal product}
f*g = \exp
(\frac{i\hbar}{2}(\partial_{\xi}\partial_{y}-\partial_{\eta}\partial_{x}))f(x,
\xi)g(y, \eta) |_{ x=y,\xi=\eta}
\end{equation}
(the Moyal product). This is a deformation quantization of $\bbR ^{2n}$.
\end{example}
We will denote by $W$ (the Weyl algebra) the algebra
\newline$\Cxxi$ with the Moyal product \eqref{eq:Moyal product} (we always
denote formal variables by $\fx$, $\fxi$). One can identify $W$
with the ring of operators of the form $\sum A_{\alpha \beta}\fx
^{\alpha}(i \hbar \frac{ \partial }{\partial \fx})^{\beta}$ on
$\bbC[[\fx,\hbar]]$. This identification takes $\fx ^{\alpha}\xi
^{\beta}$ to the symmetrized product $\fx ^{\alpha}(i\hbar
\frac{\partial}{\partial \fx}) ^{\beta}$ (the Weyl
identification). Note that, if one puts
\begin{equation} \label{eq:grading of W}
|\fx ^k| = |\fxi ^k| = 1 ,\, |\hbar| = 2,
\end{equation}
then $W$ becomes a direct product of its graded components
$$W = \prod _{k \geq 0} W _k$$
Put
\begin{equation} \label{eq:Lie algebra g}
{\g}  = \frac{1}{i\hbar}W / \frac{1}{i\hbar} \bbC [[\hbar]]
\end{equation}
with the bracket $a*b-b*a$. This Lie algebra is isomorphic to the algebra of
continuous derivations of $W$ via $a \mapsto \operatorname{ad}(a)$. The Lie
algebra $\g$ splits into the product of its graded components
$$\g = \prod _{k \geq -1} \g _k$$
Note that $\g_{-1} = \frac{1}{i\hbar}\bbC ^{2n}$ and $\g _0 =
{\frak {sp}} (2n, \bbC)$. Also, the group $\SP (2n, \bbC)$ acts on
$W$ by linear changes of coordinates. This action preserves the
product. Its infinitesimal action coincides with the adjoint
action of $\g _0$.

Put
\begin{equation} \label{eq:Lie algebra tg}
\tg = \frac{1}{i\hbar}W
\end{equation}
with the bracket $a*b-b*a$. One has
$$\tg = \prod _{k \geq -2} \tg _k$$ where
$\tg _{-2} = \frac{1}{i\hbar}\bbC,$ $\tg_ {-1} = \frac{1}{i\hbar}\bbC ^{2n}$
and
$$\tg _0 = {\frak{sp}} (2n, \bbC) \oplus \bbC$$
canonically. The subalgebra ${\frak{sp}} (2n, \bbC)$ is formed by
$\frac{1}{i\hbar} q(\fx, \fxi)$ where $q$ are quadratic functions.
\subsection{Fedosov connections} \label{subsection:Fedosov} For any
symplectic manifold $M$, we form a bundle of associative algebras $W = W_M$
and the bundles of Lie algebras $\g = \g _M$, $\tg = \tg _M$. We define them
to be the bundles associated to the $\SP$-equivariant algebras $W$, $\g$,
etc. {\em A Fedosov connection} is an operator
$$\nabla: \Omega ^{\bullet}(M, W) \to \Omega ^{\bullet+ 1}(M, W)$$
such that:
\begin{enumerate}
\item
$$\nabla =  A_{-1} + \nabla _0 + A_ 1 + \ldots$$
where $A_k \in \Omega ^{1}(M, \g _k)$ and $\nabla _0$ is a connection in
$TM$ preserving $\omega$;
\item
$$A_{-1} \in \Omega ^{1}(M, \g _{-1})= \Omega ^{1}(M, T^*_M)$$
is minus the map $T_M \to T^*_M$ defined by $\omega$;
\item
$\nabla ^{2}$ = 0
\end{enumerate}
{\em A lifting} of $\nabla$ is an expression
$$\widetilde{\nabla} =  \tA _{-1} + \nabla _0 + \tA _0 + \tA _ 1 +
\ldots$$
where $\nabla _0$ is the same connection which is now viewed as $\tg
_0$-valued, $\tA _k \in \Omega ^{1}(M, \tg _k)$, and the image of $\tA _k$
under the map induced by the projection $\tg \to \g$ is $A_k$. (For the sake
of uniformity we put $A_0 = 0$).

The following theorem is essentially due to Fedosov \cite{Fe}. For expositions closer to ours, cf. \cite{BNT}, \cite{NT1}. Another approach, which is valid for algebraic varieties, is contained in \cite{BK}.

\begin{thm} \label{thm:classification of deformations}
1) For any Fedosov connection $\nabla$,
$$\bbA _M = \bbA_M ^{\nabla}=\operatorname{ker}(\nabla:
C^{\infty} (M,W) \to \Omega ^{1}(M, W))$$
is an algebra which is isomorphic
to $C^{\infty} (M)[[\hbar]]$ as a $\bbC [[\hbar]]$-module. The resulting
product on $C^{\infty} (M)[[\hbar]]$ is a deformation quantization.

2) Any $\nabla$ admits a lifting $\widetilde{\nabla}$, and
\begin{equation} \label{eq:Weyl curvature}
{\widetilde{\nabla}} ^{2} = \theta = \frac{1}{i\hbar} \omega + \sum _{k=0}
^{\infty}(i\hbar)^k \theta _k
\end{equation}
where $\theta _k $ are closed two-forms. For two different liftings of $\nabla$, the forms $\theta$ are cohomologous.

3) Deformation quantizations $\operatorname{ker}(\nabla)$ and
$\operatorname{ker}(\nabla ')$ are isomorphic if and only if the curvatures
of their liftings are cohomologous.

4) Given two lifted Fedosov connections with the same curvature form $\theta$, there is an isomorphism between $\operatorname{ker}(\nabla)$ and
$\operatorname{ker}(\nabla ')$ which is canonical up to a canonical inner isomorphism.

5) Any deformation quantization is isomorphic to
$\operatorname{ker}(\nabla)$ for some $\nabla$.

6) For any closed $\theta$ such as in \eqref{eq:Weyl curvature}, there is a
Fedosov connection $\nabla$ with a lifting $\tn$ such that $\tn ^{2} =
\theta$.
\end{thm}

\subsection{Groups of automorphisms of $W$ and gauge transformations} \label{ss:groupoids}
Put $\tg _{\geq 1} = \prod _{k \geq 1}\tg _{k}$. This is a pronilpotent Lie
algebra, so one can define
$$\tG _{\geq 1} = \exp \tg _{\geq 1}$$
Put also
\begin{equation} \label{eq:definition of bGN}
{\overline G} _{\geq 0} = \SP (2n, \R) \ltimes \tG _{\geq 1}
\end{equation}
\begin{equation} \label{eq:definition of tGN}
\tG _{\geq 0}^N = \TSPN (2n, \R) \ltimes \tG _{\geq 1}
\end{equation}
Note that $\tG _{\geq 1}$ acts on $\tg$-valued connections by
gauge transformations. One can show that
\begin{lemma} \label{lemma:gauge equivalence} Two Fedosov connections
$\nabla$, $\nabla '$ define isomorphic star products if and only if they are
gauge equivalent, if and only if they have gauge equivalent liftings. Two lifted Fedosov connections are gauge equivalent if and only if their curvature forms are equal.
\end{lemma}
This is the key part of the proof of the statements 3 and 4 of theorem
\ref{thm:classification of deformations}. For example, to prove 4, observe that a gauge equivalence defines an isomorphism of corresponding bundles with connection, hence of the algebras of horizontal elements; two gauge equivalences of lifted connections differ by a gauge auto-equivalence, which is by definition an invertible section of the bundle $W$ which is horizontal under $\nabla$, hence an invertible element of $\operatorname{ker}(\nabla)$.
\subsubsection{The Weil representation} \label{sssss:weil} One can extend the Weil representation (cf. \ref{sss:weil}) as follows. Introduce a filtration on $\bbC[[\fx ^1, \ldots, \fx ^n, \hbar, \hbar ^{-1}]$ which is multiplicative, $\fx ^k$ are in $F^1$, and $\hbar $ in $F^2$. Let
$${\widehat{V}}^{\operatorname{Weil}} = \bigoplus _{T} \exp(\frac{iT\fx ^2}{2\hbar}){\widehat{{\bbC}}[[\fx^1, \ldots, \fx^n, \hbar, \hbar ^{-1}]},$$
where the completion on the right is with respect to the filtration $F$ and the summation is taken over all symmetric complex $n \times n$ matrices $T$ with positive definite imaginary part. Let ${\widehat{V}}_0 ^{\operatorname{Weil}}$ be a similar sum, but taken over all real $n\times n$ symmetric matrices. The action of $\TSP$ on ${{V}}^{\operatorname{Weil}}$ extends to an action of ${\widetilde{G}}_{\geq 0}$ on ${\widehat{V}}^{\operatorname{Weil}}$. Similarly, the partial action  of $\TSP$ on ${{V}}_0 ^{\operatorname{Weil}}$ extends to a partial action of ${\widetilde{G}}_{\geq 0}$ on ${\widehat{V}}_0 ^{\operatorname{Weil}}$.

We treat ${\widetilde{G}}_{\geq 0}$ as a Lie group whose Lie algebra is ${\widetilde{\g}}_{\geq 0}/{\bbC}$.
\begin{lemma} \label{lemma:estabilizadores}
The subgroup $P$ of ${\widetilde{G}}_{\geq 0}$ preserving the subspace $V_{T=0}={\bbC}[[\fx, \hbar]]$ is the Lie group of the Lie subalgebra $\{\fih f | f \in \fxi W_{\geq 1}+\hbar W_{\geq 1}$\}.

The subgroup $N$ of $P$ of those elements whose action on the subspace $V_{T=0}={\bbC}[[\fx, \hbar]]$ is identity modulo $\hbar$ is the Lie group of the Lie subalgebra $\{\fih f | f \in \fxi ^2 W_{\geq 0}+\hbar \fxi W_{\geq -1} + \hbar ^2 W\} $.
\end{lemma}
\subsection{Fedosov construction and Lagrangian submanifolds}
\label{ss:Fedosov construction and Lagrangian submanifolds} Let
$L$ be a Lagrangian submanifold of a symplectic manifold $M$. We
call a Fedosov connection $\nabla$ compatible with $L$ if the
restriction of $\nabla$ to $L$ preserves the left ideal
$WT_L^{\perp}$ where $T_L^{\perp} \subset T^*_M \subset W$ is the
annihilator of $T_L$.

The group of gauge transformations $\exp \tg _{\geq 1}(L)$ acts on
such connections, where
\begin{equation} \label {eq:Lie algebra g(L)}
\tg _{\geq 1}(L) = \{ \sigma \in \Gamma(M, \tg _{\geq 1}) \;:\; \sigma |_L \in \foh WT_L^{\perp} \}
\end{equation}
The following is, essentially, a particular case of the statements contained in \cite{Bo}, \cite{W}.

\begin{lemma} \label{lemma:gauge equivalence with L} The map
sending $\nabla$ to the cohomology class of ${\widetilde
{\nabla}}^2$ defines a bijection between the set of gauge
equivalence classes of Fedosov connections compatible with $L$ and
the affine set
$$\foh [\omega ] + H^2 (M,L)[[\hbar]]$$
\end{lemma}
The proof goes exactly as for usual Fedosov connections.
\subsection{The jet bundle $J_M$}          \label{ss:the jet bundle}
For any manifold $M$ of dimension $m$, let $J$ be the bundle of
jets of $C^ {\infty}$ functions. If $ M = \cup U _{\alpha}$ is an
open cover and a coordinate system $x^{1}, \ldots, x^m$ is chosen
on every $U_ {\alpha}$, then we identify $J|U_ {\alpha}$ with $U_
{\alpha} \times \bbC [[\fx]]$ where we denote by $\bbC[[\fx]]$ the
algebra $\bbC [[\fx ^{1}, \ldots, \fx ^m]]$. The transition
functions of the bundle $J$ are
$$G_{\alpha \beta}: U _{\alpha} \cap U _{\beta } \to \operatorname{Aut}
\bbC[[\fx]]$$ defined as follows:
\begin{equation} \label{eq:transition functions of jets}
G_{\alpha \beta}(x): \fx \mapsto g_{\alpha \beta} (\phi
_{\beta}(x) + \fx) - g_{\alpha \beta} (\phi_{\beta}(x))
\end{equation}
where
$$\phi _{\alpha}: U_{\alpha} \hookrightarrow \bbR ^N$$
are the coordinate embeddings and
$$g_{\alpha \beta} = \phi _{\alpha} \phi _{\beta} ^{-1}$$
The jet bundle is filtered by powers of the ideal $(\fx ^{1},
\ldots, \fx ^m)$, and the associated graded bundle of algebras is
$S[T^*_M]$, the symmetric algebra of the cotangent bundle. Using the fact that $C_M ^{\infty}$ is an acyclic sheaf, one
shows that, noncanonically,
\begin{equation} \label{eq:isomorphism between jets and its gr}
J_M \isomoto S[[T^*_M]]
\end{equation}
(the completion of $S[T^*_M]$) as bundles of algebras.

If a deformation quantization is given on $M$, then $J_M[[\hbar]]$
becomes a bundle of algebras. Locally, using the grading as in
\eqref{eq:grading of W}, put
$$F^k J[[\hbar]] = \prod _{p \geq k}J[[\hbar]]_p$$
The transition functions and the product preserve this filtration.
The completed associated graded bundle of algebras is the Weyl
bundle $W$. As above, one can show that
\begin{equation} \label{eq:isomorphism between jets and Weyl}
J_M[[\hbar]] \isomoto W_M
\end{equation}
as bundles of algebras.

For any $M$ there is the canonical flat connection
\begin{equation} \label{eq:canonical connection}
\nabla _{\operatorname{can}}: \Omega ^{\bullet}(M, J) \to \Omega
^{\bullet + 1}(M, J)
\end{equation}
which preserves the product. In coordinates,
$$ (\nabla _{\operatorname{can}}f)(x,\fx) = \sum _{k=1}^m (\frac{\partial f
}{\partial {x^k}} - \frac{\partial f }{\partial {\fx ^k}}) dx ^k$$
The kernel of $\nabla _{\operatorname{can}} | C^ {\infty}(M,J)$ is
canonically isomorphic to the algebra $ C^ {\infty}(M)$.

If a deformation quantization is given on $M$, the image of
$\nabla _{\operatorname{can}}$ under the isomorphism
\eqref{eq:isomorphism between jets and Weyl} becomes a Fedosov
connection. This proves the assertion 5) of theorem
\ref{thm:classification of deformations}.
\subsection{The groupoid $\tG$} \label{ss:groupoid tG}
This subsection is not used in the rest of the paper. We include it because we feel that its content might be useful in a modified theory of modules over deformation quantization algebras which is more suitable for applications.

Let $M$ be a symplectic manifold with an
$\TSPN$-structure. Let $\nabla$ be a Fedosov connection with a lifting
$\widetilde{\nabla}$, and let $\bbA _M = \operatorname{ker} \nabla $ is the
deformed sheaf of algebras of smooth functions. For two open subsets $U$ and
$V$ of $M$, let
$$G_{UV} = \operatorname{Iso} (\bbA_U, \bbA_V)$$
(the set of continuous isomorphisms of algebras). These sets form a groupoid
whose objects are open subsets for $M$. In this subsection we define the
groupoid $\tG _{UV}$ together with a surjection
\begin{equation} \label{eq:tGNM}
\tG _{UV}  \to G _{UV}
\end{equation}
such that there are central extensions
\begin{equation} \label{eq:tGNM 1}
0 \to \bbZ / N \to \tG _{UU}  \to G _{UU} \to 1
\end{equation}
Observe that $\SP (2n,
\R)$ acts on the group $\tG _{\geq 1}$ by conjugations, therefore we can construct a bundle of groups on $M$. If an $\TSPN
(2n, \R)$-structure on $M$ is given, then one defines the induced
groupoid $\tG _{\geq 0}^N$ with the manifold of objects $M$.
Locally in coordinates, for $x,y \in M$, $\tG _{\geq 0}^N (x,y) =
\tG _{\geq 0}^N$ (cf. \eqref{eq:definition of bGN}); the transition functions are left and right
multiplications by the $\TSPN$-valued lifted transition functions
of the tangent bundle $T_M$.

To define an element of $\tG _{UV}$, start with a symplectomorphism $f: V
\isomoto U$. Let $\widetilde{g} (x) \in \tG_{\geq 0} ^N (fx, x)$, $x\in V$,
be a smooth family. We require the induced family $g(x): W_{fx} \isomoto
W_x$ to preserve the Fedosov connection. By $\tG _{UV}$ we denote the set of
all such families $\widetilde{g} (x)$.

This construction is an extension of a similar construction for
symplectomorphisms which was defined in \cite{Se}.
\subsection{Modified Fedosov construction} \label{4.7} In this subsection, we
observe that the Fedosov construction can be extended as follows.
Recall that Fedosov's Weyl bundle $W_M$ is the bundle whose fiber
is the Weyl algebra $W$ and whose transition functions are the
images of the transition functions $g^T_{\alpha \beta} \in \SP
(2n)$ in the group $G_{\geq 0}$ of automorphisms of $W$. A Fedosov
connection is a flat connection of a special kind on this bundle. To define
its {\em lifting}, we used the fact that the transition functions of the
bundle $W$ admit a lifting to ${\overline G}_{\geq 0} = \SP (2n,
\R) \ltimes \tG _{\geq 1}$.

Now, let us start with any bundle of algebras ${\cal W}$ whose
transition functions take values in $G_{\geq 0}$:
$$G_{\alpha \beta} : U_{\alpha} \cap U_{\beta} \to G_{\geq 0}$$
We require that the projection of $G_{\alpha \beta}$ from $G_{\geq
0}$ to $\SP (2n)$ coincide with $g^T_{\alpha \beta}$. A lifting of
the transition functions is by definition a ${\overline G}_{\geq
0}$-valued \v{C}ech one-cocycle ${\tG} _{\alpha \beta}$ whose
image under the projection ${\overline G}_{\geq 0} \to {G}_{\geq
0}$ is $G _{\alpha \beta}$.

By definition, a Fedosov connection in ${\cal W}$ is a flat
connection which preserves multiplication and whose $A_{-1}$ term
is as in \ref{subsection:Fedosov}. Given a lifting of the
transition functions of ${\cal W}$, define a lifting of a Fedosov
connection $\nabla$ to be a $\tg$-valued connection
$\widetilde{\nabla}$ whose image under the projection $\tg \to \g$
is $\nabla$. More explicitly, it is a collection of forms
$${\widetilde {A}}_{\alpha} = \sum _{k = -1}^{\infty} {\widetilde
{A}}_{\alpha , k};\;\; {\widetilde {A}}_{\alpha , k} \in \Omega ^1
(U_{\alpha}, \tg _k)$$ such that
$${\widetilde {A}}_{\alpha} = {\operatorname{Ad}}(\tG _{\alpha
\beta})({\widetilde {A}}_{\beta}) - {\tG} _{\alpha \beta} ^{-1}
d{\tG} _{\alpha \beta}$$
The following is a straightforward generalization of \ref{thm:classification of deformations}.
\begin{thm} \label{thm:modified Fedosov construction}
1) For any ${\cal W}$, there exist a Fedosov connection $\nabla$, a
lifting ${\tG} _{\alpha \beta}$ of the transition functions, and a
lifting ${\widetilde {\nabla}}$ of $\nabla$. The algebra
${\operatorname {ker}}(\nabla :\Omega ^0 (M, {\cal{W}}) \to \Omega
^1 (M, {\cal{W}}))$ is isomorphic to a deformation quantization of
$(M, \omega)$. 

2) For two bundles $\cal W$ and ${\cal W}'$ with lifted transition functions $\widetilde {G}_{\alpha \beta}$ and $\widetilde {G'}_{\alpha \beta}$ and for two lifted Fedosov connections $\widetilde {\nabla}$ and $\widetilde {\nabla '}$, if $\widetilde {\nabla} ^2 =( \widetilde {\nabla '}) ^2$ then the algebras $\operatorname{ker} (\nabla)$ and $\operatorname{ker}(\nabla ')$ are isomorphic. Moreover, the isomorphism is canonical up to a canonical inner automorphism.

3) The curvature form
$$\theta = {\foh}\omega + \sum_{k=0}^{\infty} (i\hbar )^k
\theta_k, \;\; \theta _k \in \Omega ^2 (M),$$ is closed. Its
cohomology class is the complete invariant of the deformation up
to isomorphism.
\end{thm}
\begin{remark} \label{rmk:lifted bundles of jets} In the construction
above, one can take $G _{\alpha \beta}$ to be the transition
functions of the bundle of jets, and $\nabla$ to be the canonical
connection $\nabla _{\operatorname {can}}$. We will see an example
of this in \ref{ss:Differential operators and the deformation
quantization}.
\end{remark}

\subsection{The canonical stack on a symplectic manifold} 

We can \newline
strengthen the statement of Theorem \ref{thm:modified Fedosov construction} as follows.
\begin{prop} \label{prop:stack of connections}
1) For two bundles $\cal W$ and ${\cal W}'$ with lifted transition functions $\widetilde {G}_{\alpha \beta}$ and $\widetilde {G'}_{\alpha \beta}$ and for two lifted Fedosov connections $\widetilde {\nabla}$ and $\widetilde {\nabla '}$, if $\widetilde {\nabla} ^2 =( \widetilde {\nabla '}) ^2$ then there is a canonical isomorphism of algebras $G(\tnabla, \tnabla '): \operatorname{ker} (\nabla') \to \operatorname{ker}(\nabla )$.

2) For three bundles $\cal W$ and ${\cal W}'$ with lifted transition functions $\widetilde {G}_{\alpha \beta}$, $\widetilde {G'}_{\alpha \beta}$, $\widetilde {G''}_{\alpha \beta}$, and for three lifted Fedosov connections $\widetilde {\nabla}$, $\widetilde {\nabla '}$, $\widetilde {\nabla ''}$, if $\widetilde {\nabla} ^2 =( \widetilde {\nabla '}) ^2=( \widetilde {\nabla ''}) ^2$ then there is a canonical element $c(\tnabla, \;\tnabla ', \; \tnabla '')$ of $\operatorname{ker}(\nabla)$ which is congruent to 1 modulo $\hbar$, such that 
$$G(\tnabla, \tnabla ')G(\tnabla ', \tnabla '')=\operatorname{Ad}(c(\tnabla, \;\tnabla ', \; \tnabla '') )G(\tnabla, \tnabla '')$$

3) 
$$c(\tnabla, \;\tnabla ', \; \tnabla '')c(\tnabla, \;\tnabla '', \; \tnabla ''')=
G(\tnabla, \tnabla ')(c(\tnabla ', \;\tnabla '', \; \tnabla '''))c(\tnabla, \;\tnabla ', \; \tnabla ''')$$
\end{prop}
This provides a canonical stack of deformation quantizations on every symplectic manifold, as well as on any symplectic manifold with a pseudogroup of symplectomorphisms. Cf. \cite{Kas} and \cite{PS} for a more analytical construction which uses microdifferential operators, as well \cite{DP}  for some further discussion and applications.

{\bf Proof of the Proposition} For any $\tnabla$ and $\tnabla '$ with the same curvature form, there exists a gauge transformation $\sigma (\tnabla, \tnabla ')$ between $\tnabla '$ and $\tnabla$. Let $G(\tnabla, \tnabla ')$ be the action of this gauge transformation reduced to horizontal sections. Put also 
$$c(\tnabla, \tnabla ', \tnabla '') = \sigma(\tnabla, \tnabla ')\sigma(\tnabla ', \tnabla '')\sigma(\tnabla, \tnabla '')^{-1}$$
It is easy to see that these $G$ and $c$ satisfy all the properties stated in the Proposition.
\subsection{Differential operators and the deformation quantization of $T^*
(X)$}\label{ss:Differential operators and the deformation
quantization}
Let $X$ be a manifold. For the sheaf of rings
$\DX$ of differential operators on half-densities on $X$, let $F_p
\DX$ be the filtration by order. Let
\begin{equation} \label{eq:Rees ring}
\RDX = \bigoplus _{p\geq 0}\hbar ^p F_p \DX
\end{equation}
be the Rees ring. Let $X = \cup _{\alpha} U_{\alpha}^0$ be an open cover. A
choice of coordinates on $U _{\alpha}^0$ identifies $\RDUa$ with the ring
$\Ci_{U_{\alpha}^0}[\xi ^{1}, \ldots, \xi ^n, \hbar]$ where
$$\xi ^k = i\hbar \frac{\partial}{\partial x^k}$$
The latter ring can be identified with the ring $\Cip _{ \pi
^{-1}U_{\alpha}^0}$ where $\pi: T^*X  \to X$ is the projection and
$\Cip$ stands for the sheaf of $\Ci$ functions which are
polynomial along the fibers. We use the Weyl identification,
analogous to one that was used in the discussion after
\eqref{eq:Moyal product}.

We get
$$\phi _{\alpha}: \RDUa \isomoto \Cip _{ \pi ^{-1}U_{\alpha}^0}$$
One checks that the product on $\RDUa$ induces on the right hand side a star
product which we denote by $* _{\alpha}$. This star product extends to $\Ci
_{\pi ^{-1}U_{\alpha}}$. Furthermore, $G_{\alpha \beta} = \phi _{\alpha}
\phi _{\beta}^{-1}$ extend to isomorphisms between $* _{\alpha}$ and $*
_{\beta}$ in the sense of definition \ref{dfn:isomorphism}. Using partitions
of unity, one constructs automorphisms $T_{\alpha}$ of $* _{\alpha}$ such
that $G_{\alpha \beta} = T _{\alpha} T _{\beta}^{-1}$. This allows to define
a star product on $\TX$.

Let us recall how one identifies the above deformation in terms of
the classification theorem \ref{thm:classification of
deformations}.
\begin{prop} \label{prop:theta on T*X} The
characteristic class $\theta$ of this deformation is $\foho$
($=0$).
\end{prop}
{\bf Proof.} The statement itself is straightforward. Indeed, one checks that our deformation is isomorphic to its opposite, and it is easy to see that the characteristic class of the opposite is minus the original characteristic class. We will need, however, an explicit description of our deformation in terms of Section \ref{4.7}.

Start with an open cover $\{\pi ^{-1}(U^0 _{\alpha})
\}$ of $T^*(X)$. A choice of coordinates $x_{\alpha} = (x_{\alpha}
^1, \ldots, x_{\alpha} ^n)$ on $U^0 _{\alpha}$ determines a
coordinate system $(x_{\alpha}, \xi^{\alpha})$ on $\pi ^{-1}(U^0
_{\alpha})$. The transition functions between two different
coordinate systems are
\begin{eqnarray} \label{eq:transition functions for T*X}
x_{\alpha} & = & g_{\alpha \beta}(x_{\beta}) \nonumber \\
\xi^{\alpha} & = & ^t g_{\alpha \beta}'(x_{\beta}) ^{-1} \xi
^{\beta}
\end{eqnarray}
($^t g_{\alpha \beta}'(x_{\beta})$ stands for the transposed
Jacobi matrix).

Now consider the bundle
$\pi ^* \jets \RDX$ with the canonical connection $\pi ^* \nabla
_{\operatorname{can}}$. (To construct the bundle of jets one acts
as in \ref{ss:the jet bundle}). This is a bundle of algebras with
the fiber
\begin{equation} \label{eq:Wfin}
W_{\operatorname {fin}} = \bbC [[\fx]][\fxi, \hbar]
\end{equation}
(again, we use the Weyl identification of the two sides). Its
transition functions are given explicitly as follows:
\begin{equation} \label{eq:transition functions for lifted Rees}
\fx \mapsto   g_{\alpha \beta}(x_{\beta}
+ \fx) - x_{\alpha}
\end{equation}
\begin{equation} \label{eq:transition functions for lifted Rees 2}
\fxi  \mapsto ^t g_{\alpha \beta}'(x_{\beta} + \fx) ^{-1}
\fxi
\end{equation}
Because of the presence of the half-densities, the product in the
second equation is the commutative product, not the composition in
the Weyl algebra (as always, we use the Weyl identification
between functions and operators). The canonical connection is, in
coordinates, given by
$$\pi ^*\nabla _{\operatorname{can}}=d -\sum \frac{\partial}{\partial \fx
^k}dx^k,$$
without a
similar $\xi$ term. To correct that, apply the gauge
transformation
\begin{equation} \label{eq:gauge transformation}
\sigma _{\alpha} = \exp \operatorname{ad}\frac{1}{i\hbar}\xi
^{\alpha} \fx \in \operatorname{Aut}(W_{\operatorname {fin}})
\end{equation}
We get a new bundle of algebras whose transition functions are
\begin{equation} \label{eq:transition functions for cW}
\fx \mapsto   g_{\alpha \beta}(x_{\beta}
+ \fx) - x_{\alpha}
\end{equation}
\begin{equation} \label{eq:transition functions for cW 2}
\fxi^  \mapsto ^t g_{\alpha \beta}'(x_{\beta} + \fx) ^{-1}
(\xi^{\beta} + \fxi) - \xi ^{\alpha}
\end{equation}
Again, the multiplication in the second formula is the commutative
multiplication of power series. Therefore, these transition
functions coincide with the transition functions of the jet bundle
$J_{T^*X}$ (compare with \eqref{eq:transition functions for T*X}).
Note also that these transition functions admit a canonical
lifting to ${\overline G}_{\geq 0}$ and even to ${\widetilde
G}_{\geq 0}$. Indeed, consider the group $K$ of formal
automorphisms of the trivial line bundle on $\bbR ^n$ acting by
\begin{equation} \label{eq:definition of K}
f(\fx) \mapsto p(\fx) f(g(\fx))
|{\operatorname{det}}g'(\fx)|^{\fot}
\end{equation}
where $g$ is a formal diffeomorphism of the form $g:\fx \mapsto
a\fx + o(\fx)$, $a\in \operatorname{GL}(n,\bbR)$, and $p(\fx) \in
\bbC[[\fx]], \; p(0) = 0$. The group $K$ maps into $G_{\geq 0}$ as
follows. We can represent $W$ as an algebra of operators on $\bbC
[[\fx, \hbar]]$ by identifying, as above, $\fx ^n \fxi ^m$ with
the symmetrized product $\fx ^n (i\hbar \frac{\partial}{\partial
\fx})^m$. Then $K$ acts on these operators by conjugation. To
check that there is a canonical lifting $K \to {\widetilde
G}_{\geq 0}$, observe that
$$K = \operatorname{GL}(n,\bbR) \ltimes K_{\geq 1}$$ where $K_{\geq
1}$ is the group of elements for which $a=1$. But
$\operatorname{GL}$ maps to $\operatorname{Sp}$, and this map
lifts to $\TSP$; on the other hand, $K_{\geq 1}$ has a canonical
lifting to ${\widetilde{G}}_{\geq 1}$. Indeed,
$$K_{\geq 1}=\exp({\frak{k}}_{\geq 1})$$
where
\begin{equation} \nonumber
{\frak{k}}_{\geq 1} = \{ \foh P(\fx) \fxi + Q(\fx) | P(\fx) =
o(\fx);\; Q(0) = 0\}
\end{equation}
and this pronilpotent Lie algebra is a subalgebra of $\tg _{\geq
1}$, not just of $\g _{\geq 1}$.

As for the connection, $\pi ^*\nabla _{\operatorname{can}}$
becomes a Fedosov connection after the gauge transformation
$\sigma _{\alpha}$. It has a flat lifting which has an extra
summand $\frac{i}{\hbar}\xi dx$. Subtract it, and we get a Fedosov
connection, together with a lifting whose curvature is $\foho$.

Therefore we are exactly in the situation of Theorem
\ref{thm:modified Fedosov construction} and Remark \ref{rmk:lifted
bundles of jets}. This proves the proposition.

\section{Preliminaries from Lagrangian analysis}
\subsection{} Let us recall the classical construction from \cite{H},
\cite{GS} in terms that are suited for our purposes. Let $X$ be a
manifold. Consider an open coordinate cover $X = \cup U _{\alpha}
^{0}$, and a refinement of the cover $\pi ^{-1}U_ {\alpha}$: $\TX
= \cup U_{\beta}$; $U_{\beta}  \subset \pi ^{-1} U^{0} _{\alpha
(\beta)}$. Let $L\subset \TX$ be a Lagrangian submanifold. Note
that at this stage we do not assume any of the subsets to be
conical. Denote by ${\cal E}_L$ the local system with the
transition functions $\exp (\frac{i \pi}{2} c)$ where $c$ is a
$\bbZ$-valued one-cocycle representing the Maslov class of $L$.

1) For any section
$$a \in \Gamma _c (\Ub , |\Omega _L| ^{\frac {1}{2}} \otimes {\cal E}_L)$$
and any $\hbar \neq 0$, one can construct a half-density

\begin{equation} \label{eq:ahat}
\widehat{a}_{\beta, \hbar} \in \Gamma  (\Uoab , |\Omega _X| ^{\frac
{1}{2}})
\end{equation}

2) For any smooth function $f$ on $\pi ^{-1} (\Uoa)$, polynomial along the
fibers, and any $\hbar \neq 0$, one can construct a differential operator $\widehat {f}
_{\alpha, \hbar}$ on half-densities on $\Uoa$.

3) If $a$ is supported in $\Ub \cap \Ubo$ then
$$ \aohat = \ahat + \sum _{k = 1}^{\infty} (i\hbar) ^k \Rkhat _{\beta,
\hbar} + \Ohi $$ as $\hbar \to 0$, where $\Rk$ are differential operators on sections of
$|\Omega _L| ^{\frac {1}{2}} \otimes {\cal E}_L)$.

4) If $f$ is supported in $\Uoa \cap U^{0} _{\alpha _1}$, then
$$\fohat = \fhat + \sum _{k = 1}^{\infty} (i\hbar) ^k \Tkhat _{\alpha,
\hbar} +\Ohi
$$
where $\Tk$ are differential operators.

5)
$${\widehat{f}} _{\alpha (\beta), \hbar} ({ \widehat{a}}_{\beta, \hbar}) =
(f|_{L} \cdot a)_{\beta, \hbar} + \sum _{k = 1}^{\infty} (i\hbar) ^k \Pkhat
_{\beta, \hbar} + \Ohi$$
where $\Pk$ are bidifferential expressions depending on $a$ and on the jet
of $f$ at $L$.

6)
$${\widehat{f_1}} _{\alpha, \hbar} \circ {\widehat{f_2}} _{\alpha, \hbar} =
{\widehat{f_1 \cdot f_2}} _{\alpha, \hbar} + \sum _{k = 1}^{\infty} (i\hbar)
^k \Dkhat _{\alpha, \hbar} + \Ohi$$
where $\Dk$ are bidifferential expressions.

We see that the asymptotic expressions 3) - 6) define a
deformation quantization of $\TX$ and a sheaf of modules $\VHL$
over the deformed algebra ${\bbA} _{\TX}= \Ci _{\TX}[[\hbar]]$,
supported on $L$.

Let us briefly recall how to construct $\ahat$. Locally on $\Ub$,
consider {\em a phase function} of $L$: if $x = (x^{1},
\ldots,x^{n} )$ are coordinates on $U_{\alpha (\beta)} \subset X$,
let $\theta$ be a variable in $\bbR ^k$; a phase function is a
function $\varphi (x,\theta)$ such that:

i)
$$L = \{(x,\xi) | d_{\theta} \varphi (x,\theta) = 0; \, \xi = d_{x} \varphi
(x,\theta) \}$$
ii) The Hessian $(\varphi _{x \theta}, \, \varphi _{\theta \theta})$ is of
maximum rank $k$.

Given a phase function, the map
$$
i: \{(x,\theta) | d_{\theta} \varphi (x,\theta) = 0\} \to L;
$$
$$
\, (x,\theta) \mapsto  (x, d_{x} \varphi (x,\theta))
$$
is a local diffeomorphism. The left hand side is a submanifold in $\bbR
^{n+k}$.

The function $\varphi (i ^{-1}x)$, which we still denote by $\varphi$,
satisfies $d\varphi = \xi dx$, the right hand side being the canonical
one-form on $\TX$ (its differential is $\omega$, so it is closed on $L$).

The H\"{o}rmander construction is as follows. Start with a local
section $a$ on $\Ub$. Given a phase function $\varphi = \varphi
_{\beta}$ on $\Ub$, and given local coordinates on $U_{\alpha
(\beta)}$, represent, locally, $a$ as a function on $\{(x,\theta)
| d_{\theta} \varphi (x,\theta) = 0\}$. Extend it to a function of
$x, \theta$ which is zero away from a small neighborhood of $i
^{-1}L$ as follows. Subdivide the $n+k$ variables $(x,\theta)$
into two groups $y\in \bbR ^n$, $z \in \bbR ^k$, such that the
Hessian $k\times k$ matrix ${\partial}^2 _{z\theta}\varphi$ is
nondegenerate. Observe that the restriction of the map
\eqref{eq:map to L} to the subspace $z=0$ is a local
diffeomorphism with $L$; extend $a$ to a function in $y$ only, and
multiply by a function in $y$ which is zero away from a
neighborhood of the origin. Now define
\begin{equation}
\label{eq:definition of a(beta)} \ahat = [\int e ^{\frac
{i}{\hbar} \varphi _{\beta}(x, \theta)} a(x,\theta)d\theta]|
dx|^{\fot}
\end{equation}
One can now proceed to generalize this to the case when $L$ is a homogeneous
Lagrangian submanifold, $\varphi$ is homogeneous of degree one in $\theta$,
and $a(x,\theta)$ satisfying certain standard growth conditions with respect
to $\theta$. One can still define $\ahat$ as distributional half-densities,
whose action on a test half-density $u(x)| dx|^{\fot}$ is defined as
$$\ahat (u)= \int e ^{\frac {i}{\hbar} \varphi _{\beta}(x, \theta)}
a(x,\theta)u(x) d\theta dx$$
The latter integral is taken using the stationary phase method as explained
in \cite{GS} and \cite{H}.

\section{Lagrangian analysis and deformation quantization} \label{s:Hormander in deformation quantization}
\subsection{}Let $L \subset \TX$ be a Lagrangian submanifold. As
above, let $X= \cup _{\alpha} \Uoa$ and $\TX = \cup _{\beta} \Ub$,
a cover which is a refinement of $ \{\pi ^{-1}\Uoa\}$. Let
$\varphi _{\beta}$ be a phase function of $L|_{\Ub}$.Using these
data, we will construct a sheaf of modules over $\bbA
^{\hbar}_{\TX} \otimes _{\bbC[[\hbar]]}\bbK$ where $\bbA
^{\hbar}_{\TX} $ is the deformation quantization discussed in
\ref{ss:Differential operators and the deformation quantization}
and $\bbK = \bbC [[\hbar]][\hbar ^{-1}, e ^{\frac{i}{\hbar}a }|a
\in \bbR]$.

Denote by $V_{\beta}$ the space of formal expressions
\begin{equation} \label{eq:definition of local module V}
\frac{1}{({2 \pi \hbar})^{k/2} }[\int e^{\fih \varphi (x, \theta)}
a(x, \theta) d\theta] |dx|^{\fot}
\end{equation}
where $a(x,\theta)$ is a $\bbC[[\hbar]]$-valued smooth function
and $k$ is the dimension of the $\theta$ space (we will mostly use local phase functions of special kind for which $k=n$). More precisely,
\begin{equation} \label{eq:definition of local module V 1}
V _{\beta} = \{ a(x, \theta)) \} / \eq
\end{equation}
where
\begin{equation} \label{eq:definition of local module V 2}
\varphi _{\theta} \cdot a \eq i\hbar a_{\theta}
\end{equation}
and $a(x,\theta)$ is a $\bbC[[\hbar]]$-valued smooth function on
the preimage of $\Ub$ under the map
\begin{equation}\label{eq:map to L}
(x, \theta)
\mapsto (x, d_x \varphi (x, \theta))
\end{equation}
Note that $V_{\beta}$ as a $\bbC [[\hbar]]$-module is isomorphic
to $\Ci (L \cap \Ub)[[\hbar]]$. Indeed, using the discussion
before the formula \eqref{eq:definition of a(beta)}, we see that
the map
$$a(y) \mapsto
(a \operatorname{mod} \eq) \in V_{\beta}$$ is an isomorphism.

Now we have to define the transition functions
$$G_{\beta \gamma}: V_{\beta} | U_{\beta } \cap U_{\gamma}
\isomoto V_{\gamma } | U_{\beta } \cap U_{\gamma}$$
and the action
of $\pi ^* \RDUa $ on $V_{\beta}$ where $\alpha = \alpha (\beta)$.
Both are suggested by \eqref{eq:definition of local module V}. The
action is defined by
\begin{eqnarray} \label{eq:action of A on V}
x^m \cdot a  & = & x^m a \nonumber \\
\xi ^m \cdot a & = & i\hbar \frac{\partial a}{\partial x^m} -
\frac{\partial \varphi}{\partial x^m} \cdot a
\end{eqnarray}
for $m = 1, \ldots, n$. This action preserves the equivalence
relation \eqref{eq:definition of local module V 2}.

As for the transition functions, let us start, following
H\"{o}rmander, by introducing coordinate changes
\begin{equation} \label{eq:coordinate change in a phase function}
\varphi(x, \theta) \mapsto \varphi(g(x), \rho(x,\theta))
\end{equation}
where $g$ is a local diffeomorphism. It is straightforward that a phase function obtained by a coordinate change from $\varphi$ defines the same Lagrangian submanifold as $\varphi$. H\"{o}rmander proved that,
locally on $U_{\beta}$, any two phase functions differ by such a
coordinate change, followed by addition of $\pm \sum
_{i=1}^{c}\theta_i^2/2$. As we will see later, the numbers $\pm c
= \mu_{\beta \gamma}$ define a cocycle representing twice the Maslov
class).

Let us write down the transition functions corresponding to the
change \eqref{eq:coordinate change in a phase function} where
$$g = g_{\beta \gamma} = g_{\alpha(\beta) \alpha(\gamma)}$$
are the transition functions of the manifold $X$ and $\rho = \rho
_{\beta \gamma}$. They act on the equivalence classes of formal
expressions \eqref{eq:definition of local module V} as follows:
\begin{eqnarray} \label{eq:transition functions, 1}
G_{\beta \gamma}:\frac{1}{({2 \pi \hbar})^{k/2} }\int e^{\fih{\varphi
(x,\theta)}}a(x_{\beta},\theta)d\theta
|dx_{\beta}|^{\fot} \mapsto \nonumber \\
\frac{1}{({2 \pi \hbar})^{k/2} }\int e^{\fih\varphi (g_{\beta
\gamma}(x_{\gamma}), \rho_{\beta \gamma}(x_\gamma,
\theta))}a(g_{\beta \gamma}(x_{\gamma}),
\rho_{\beta \gamma}(x_\gamma, \theta))\times \\
\times|\operatorname{det} \frac{\partial \rho_{\beta
\gamma}(x_\gamma, \theta))}{\partial{\theta}}|^{-1}
|\operatorname{det}g_{\beta \gamma}'(x_{\gamma})|^{\fot}d\theta
|dx_{\gamma}|^{\fot} \nonumber
\end{eqnarray}
Each summand $\pm \theta _i^2$ contributes a multiple
$$\frac{1}{({2 \pi \hbar})^{1/2} }\int e^{\pm \frac{ i\theta
^2}{2\hbar}}d\theta  = e^{\mp
\frac{i\pi}{4}}$$

To show that $G_{\beta \gamma}G_{\gamma \delta} = G_{\beta \delta}$, note that, though two phase functions may differ from one another by more than one coordinate change, two different coordinate changes define the same transformation of the space $\{a(x,\theta)\}/{\eq}$, provided that the underlying changes of the coordinate $x$ are the same.
 
\begin{example}\label{example:xi=kx} Let $L$ be given by the
equation $\xi=kx$ in $\bbR ^2$. Assume that $k\neq 0$. Consider
two phase functions of $x=x_{\beta}=x_{\gamma}$;
$$\varphi _{\beta}(x)= k\frac{x^2}{2};$$
$$\varphi _{\gamma}(x,\theta) = x\theta - k^{-1}\frac{\theta
^2}{2}$$
The phase function $\varphi _{\gamma}$ can be obtained from
$\varphi _{\beta} - \operatorname{sgn}(k) \frac{\theta ^2}{2}$ by a
coordinate change $$g_{\beta
\gamma}(x)=x;$$
$$\rho _{\beta
\gamma}(x,\theta)=\sqrt{|k|}x-\frac{\operatorname{sgn}(k)}{\sqrt{|k|}}\theta$$
The transition functions act as follows:
\begin{eqnarray} e^{\fih \frac{kx^2}{2}}a(x)|dx|^{\fot} \mapsto \nonumber
\\
\frac{1}{(2\pi)^{\fot}}e^{\operatorname{sgn}(k)\frac{\pi i}{4}}\int e^{\fih
(\frac{kx^2}{2} - \operatorname{sgn}(k)\frac{\theta ^2}{2})}a(x) d\theta
|dx|^{\fot}
\mapsto \\
\frac{1}{(2\pi)^{\fot}}\sqrt{|k|}e^{\operatorname{sgn}(k)\frac{\pi
i}{4}}\int e^{\fih (x\theta - \frac{\theta ^2}{2k})}a(x) d\theta
|dx|^{\fot}
\end{eqnarray}
In other words, an element of $V_{\beta}$ is a function of $x$ and
$\hbar$. An element of $V_{\gamma}$ is a function of $x, \theta,$
and $\hbar$, modulo equivalence
$$i\hbar \partial _{\theta}a \eq
(x-k^{-1}\theta)a$$ In every equivalence class there is unique
function depending on theta and $\hbar$ and not on $x$. The
transition function acts by taking $a(x, \hbar)$, multiplying it
by $\sqrt{|k|}e^{\operatorname{sgn}(k)\frac{\pi i}{4}}$, and then
rewriting it as a function of $\theta $ and $\hbar$ using the
above equivalence relation.
\end{example}
Let us introduce a special class of phase
functions generalizing the above example, cf. \cite{GS}. For any
$\beta$, choose coordinates on $U _{\alpha (\beta)}$ such that
$x=(x_1, x_2)$ where $x_1 \in \bbR ^{n_1}$, $x_2 \in \bbR ^{n_2}$,
$n_1 + n_2 = n$, such that $L$ can be defined by equations
\begin{eqnarray} \label{eq:parametrization of L}
\xi _1 & = & F _{x_1}(x_1, \xi _2) \nonumber \\
x_2 & = & -F_{\xi _2}(x_1, \xi _2)
\end{eqnarray}
where $\xi _1, \xi _2$ are coordinates dual to $x_1, x_2$. This is
equivalent to the requirement that the projection of $L$ to $\{\xi
_1 = x_2 =0 \}$ along $\{\xi _2= x_1 =0 \}$ is an isomorphism on
$\Ub$.

Now, put $\theta = (\xi _1, \xi _2)$ and
\begin{equation} \label{eq:definition of a phase function}
\varphi _{\beta}(x, \theta) = \varphi (x, \theta) =x_2 \xi _2 +
F(x_1, \xi _2) + \fot (\xi _1 - F_{x_1})^2
\end{equation}
This is a local phase function for $L$.

For the above phase functions, it is easy to see that $V_{\beta}$
as a $\bbC [[\hbar]]$-module is isomorphic to $\Ci (L \cap
\Ub)[[\hbar]]$. Indeed, observe that the restriction of the above
map to $\{x_1, 0, 0, \xi _2\}$ is a local diffeomorphism with $L$;
on the other hand, the map $a(x_1, \xi _2) \mapsto (a
\operatorname{mod} \eq) \in V_{\beta}$ is an isomorphism.

More precisely, the equivalence relation for $a$ is:
\begin{eqnarray} \label{eq:equivalence relation for a, special
case}
i\hbar \frac{\partial a}{\partial \xi _1} \eq (\xi _1 -
F_{x_1})a \nonumber \\
i\hbar \frac{\partial a}{\partial \xi _2} \eq (x_2 + F_{\xi _2})a
\end{eqnarray}
This equivalence allows to identify elements of $V_{\beta}$ with
functions of $x_1$, $\xi _2$, $\hbar$. Under this identification,
the algebra $\bbA ^{\hbar}_{U_{\beta}}$ acts by \begin{eqnarray}
\label{eq:action of A:special case} x_1 & \mapsto & x_1 \nonumber
\\
x_2 & \mapsto & i\hbar \frac{\partial}{\partial \xi _2} -
\frac{\partial F}{\partial \xi _2} \nonumber \\
\xi _1 & \mapsto & i\hbar \frac{\partial}{\partial x_1} +
\frac{\partial F}{\partial x_1} \\
\xi _2 & \mapsto & -\xi _2 \nonumber
\end{eqnarray}
In other words, locally,
\begin{equation} \label{eq:V as induced}
\bbA ^{\hbar}_{U_{\beta}} / I_F
\end{equation}
where $I_F$ is the left ideal generated by the local equations
$x_2 + F_{\xi _2}$, $\xi _1 - F_{x_1}$ of $L$.

Now let us describe the transition functions for this special
choice of the phase functions. This will generalize Example
\ref{example:xi=kx}.

Assume that on $\Ub$ $L$
is presented as
\begin{eqnarray} \label{eq:L on intersections 1}
\xi _1 & = & F_{x_1}(x_1, x_2, \xi _3, \xi _4) \nonumber \\
\xi _2 & = & F_{x_2}(x_1, x_2, \xi _3, \xi _4) \nonumber \\
x_3 & = & - F_{\xi _3}(x_1, x_2, \xi _3, \xi _4) \\
x_4 & = & -F_{\xi _4}(x_1, x_2, \xi _3, \xi _4)
\end{eqnarray}
and on $U _{\gamma}$
\begin{eqnarray} \label{eq:L on intersections 2}
\xi _1 & = & G_{x_1}(x_1, x_2, \xi _3, \xi _4) \nonumber \\
x _2 & = &- G_{\xi_2}(x_1, x_2, \xi _3, \xi _4) \nonumber \\
\xi_3 & = & G_{x _3}(x_1, x_2, \xi _3, \xi _4) \\
x_4 & = & -G_{\xi _4}(x_1, x_2, \xi _3, \xi _4)
\end{eqnarray}
This means that the Hessian matrix $\operatorname{Hess}_{{\xi_2},
{x_3}}(F)$ is nondegenerate. If $a_{\gamma} = G_{\beta \gamma}
(a_{\beta})$, then the following two expressions should be the same:
\begin{eqnarray}
\int e^{\fih (\xi _3 x _3 + \xi _4 x_4 + F + \fot (\xi _1 -
F_{x_1})^2 + \fot (\xi _2- F_{x_2})^2)} a_{\beta}(x_1, x_2, \xi
_3, \xi _4) d\xi _1 d\xi _2 d\xi _3 d\xi _4 = \nonumber \\
\int e^{\fih (\xi _2 x _2 + \xi _4 x_4 + G + \fot (\xi _1 -
G_{x_1})^2 + \fot (\xi _3- G_{x_3})^2)} a_{\gamma}(x_1, x_3, \xi
_2, \xi _4) d\xi _1 d\xi _2 d\xi _3 d\xi _4 \nonumber
\end{eqnarray}
Here, as above, we use the rule
\begin{equation} \label{eq:Gaussian integral}
\frac{1}{\sqrt {2 \pi \hbar}} \int e ^{\fih k x^2 /2} dx =
\frac{1}{\sqrt {ik}},
\end{equation}
$k \neq 0$, where we choose the branch of the square root
$$\sqrt {r e^{i\varphi}} = {\sqrt {r}} e ^{i\varphi /2}, \;
\varphi \neq \pi$$.

Thus,
\begin{equation} \label{eq:transition functions in terms of
Fourier}
e^{\fih G}a_{\gamma} = e^{\frac{\pi i}{4}(n_3 -
n_2)}\operatorname{Fourier}_{\xi _3 \to x_3;\; x_2 \to -\xi _2}
(e ^{\fih F} a_{\beta})
\end{equation}
Here, for $x \in \bbR ^N$,
\begin{equation} \label{eq:Fourier}
(\operatorname{Fourier} f) (\xi) = \frac{1}{ ({2 \pi
\hbar})^{N/2}} \int e ^{\fih x\xi} f(x) dx
\end{equation}
It remains to make sense of an expression
\begin{equation} \label{eq:Fourier 1}
\operatorname{Fourier} e^ {\fih F(y)}a(y)
\end{equation}
where $F$ is a smooth function with an isolated critical point
$y_0$ at which the Hessian is nondegenerate. We define this via a
well known asymptotic expansion

\begin{equation} \label{eq:statfaza}
\operatorname{Fourier} e^ {\fih F(y)}a(y) = e^ {\fih G(\eta)}
\exp (\frac{\pi i}{4}\operatorname{sgn}\operatorname{Hess}
_{y_0}F)|\operatorname{Hess}
_{y_0}F)|^{-\frac{1}{2}}\sum _{k=0}^{\infty}b_k (\eta) \hbar ^k 
\end{equation}

Here $G(\eta)$ is defined by
$$G(\eta) = \eta y + F(y),$$
$y$ being the solution of
$$\eta + F'(y) = 0$$
(the Legendre transform of $F$).

\begin{remark} \label{rmk:Fourier for formal series}
Let us stress that the above expansion makes sense for a power
series $F$ with nondegenerate Hessian and for a power series
$a(y)$. Then $G$ and $b$ are also power series. For the Fourier transform this was already explained in \ref{sss:weil} and \ref{sssss:weil}. 
\end{remark}

For open subsets $\Ub$ and $U_{\gamma}$ put
\begin{equation}\label{eq:Maslov cocycle 2 1}
c_{\beta \gamma} = \frac{1}{2}
\operatorname{signature}\operatorname{Hess}_{{\xi_2}, {x_3}}(F)
\end{equation}
By theorem \ref{thm:Maslov cocycle 2}, the cochain $c$ is a
$\frac{1}{2} \bbZ$ - valued \v{C}ech one-cocycle of $L$ which is
cohomologous to a $\bbZ$-valued cocycle representing the Maslov
class of $L$.

Let $\alpha = \xi dx$ be the canonical one-form; since $d\alpha =
\omega$, $\alpha |L$ is closed. The choice of local phase functions allows us 
to represent it by a \v{C}ech
one-cocycle $\alpha _L$ with values in $\bbR$:
\begin{equation} \label{eq:alpha}\alpha_{\beta \gamma} = \varphi _{\beta} - \varphi _{\gamma}
\end{equation}
on $L$. The right hand side is locally constant since for all $\beta$ $d\varphi _{\beta} = \alpha$ on $L$. We have proven the
following statement.
\begin{proposition} \label{prop:classical approximation of V}
The sheaf $V$ defined via local modules $V_{\beta}$ and transition
functions $G_{\beta \gamma}$ is a sheaf of $\bbA ^{\hbar} \otimes
_{\bbC [[\hbar]]} {\bbK}$-modules supported on $L$. Locally, $V
\isomoto |\Omega _L|^{\fot} \otimes _{\bbC} \bbK$, and the
transition functions are of the form
$$G_{\beta \gamma} = e^{\fih \alpha _L + \frac{\pi i}{2} \mu _L} g
_{\beta \gamma}(\hbar)$$
where $\mu _L$ is a $\bbZ$-valued
cocycle defining the Maslov class of $L$ and $g_{\beta \gamma} = 1
(\operatorname {mod} \hbar)$
\end{proposition}
\begin{definition} \label{dfn:Hormander module}
We denote the above sheaf of modules by $V _L$.
\end{definition}

\section{The Lagrangian jet bundle}
\subsection{} Next we observe that $V_L$ is the sheaf of horizontal
sections of a module over the algebra of jets of functions on
$\TX$ with the star product constructed above. This module of jets
will be equipped with a flat connection compatible with the
canonical connection in the bundle of jet algebras.

First, define, for an open subset $\Ub$ and a phase function
$\varphi = \varphi _{\beta}$,
\begin{equation} \label{eq:module J}
J_{\beta} = \{e^{\fih \varphi _{\beta}(x+\fx, \theta + \ftheta)}
a(x, \theta;\fx, \ftheta; \hbar)|dx|^{\fot}\}/ \eq
\end{equation}
where $(x,\theta)$ are in the preimage of $L \cap \Ub$ under the
map $(x,\theta) \mapsto (x, d_x \varphi (x, \theta))$ and
\begin{equation} \label{eq:eqivalence for jet module}
e^{\fih \varphi (x+\fx, \theta + \ftheta)} \varphi _{\theta}
(x+\fx, \theta + \ftheta)a|dx|^{\fot} \eq i\hbar e^{\fih \varphi
(x+\fx, \theta + \ftheta)} a _{\ftheta} |dx|^{\fot}
\end{equation}
Locally, $J_{\beta}$ is isomorphic to the space of sections on
$\Ub$ of the bundle of jets of half-densities on $L$ ($\bbC
[[\hbar]]$-valued).

One defines the transition functions $J_{\beta} \isomoto
J_{\gamma}$ on $\Ub \cap \Ug$ and the action of $\pi ^*
\operatorname{jets} \RDX $. The latter is defined by
\begin{eqnarray}
& \fxi ^m: e^{\fih \varphi}a |dx|^{\fot} \mapsto e^{\fih
\varphi}(i\hbar
\partial _{\fx ^m}a - \varphi _{\fx ^m}a)|dx|^{\fot} \nonumber \\
&\fx ^m : e^{\fih \varphi}a |dx|^{\fot} \mapsto e^{\fih \varphi}
\fx ^m a |dx|^{\fot} \nonumber
\end{eqnarray}

The transition functions corresponding to a change
\eqref{eq:definition of local module V} act by
\begin{eqnarray} \label{eq:eqivalence for jet module 1}
&a(x_{\beta}, \theta) \mapsto \\
&a(g_{\beta \gamma}(x_{\gamma}),
\rho_{\beta \gamma}(x_\gamma, \theta); g_{\beta \gamma}(x_{\gamma}
+ \fx)-x_{\beta},
\rho_{\beta \gamma}(x_\gamma + \fx, \theta + {\widehat{\theta}})-\rho_{\beta
\gamma}(x_\gamma, \theta))\times \nonumber \\
&\times|\operatorname{det} \frac{\partial \rho_{\beta
\gamma}(x_\gamma + \fx, \theta +
\widehat{\theta}))}{\partial{\theta}}|^{-1}
|\operatorname{det}g_{\beta \gamma}'(x_{\gamma} + \fx)|^{\fot}
\nonumber
\end{eqnarray}

One defines also the flat connection
$$\nabla = (\frac{\partial}{\partial x} -  \frac{\partial}{\partial \fx
}) dx = \sum (\frac{\partial}{\partial x^m } -
\frac{\partial}{\partial \fx ^m }) dx^m$$

One gets a bundle of $\pi ^* \operatorname{jets} \RDX$-modules
with a flat connection which is compatible with the canonical
connection on the jet bundle. We would like to modify this
construction as follows. Recall that we have constructed in the
proof of proposition \ref{prop:theta on T*X} a multiplication on
the jet algebra $J_{T^*X}[[\hbar]]$ and a Fedosov connection whose
lifting has the curvature $\foh \omega$. We denote the resulting
bundle of algebras by ${\cal{W}}_{\TX}$. We would like to get a
bundle of ${\cal{W}}_{\TX}$-modules with a flat connection which
is compatible with the Fedosov connection. To achieve that, we
have to modify our module $J$.
\begin{definition} \label{dfn:Hormander jets}
For a point $\bf x$ of $L \cap U_{\beta}$ represented by
$(x,\theta)$, put
\begin{equation} \label{eq:new J}
(J^H_L)_{\bf x} = \{e^{\fih \varphi_{\beta}(x,\theta; \fx,
\ftheta) }a(x, \theta; \fx, \ftheta; \hbar) \} / \eq
\end{equation}
where
\begin{equation} \label{eq:phase function at a point}
\varphi _{\beta}(x,\theta; \fx, \ftheta)=\varphi_{\beta}(x + \fx,
\theta + \ftheta)-\varphi_{\beta}(x,\theta) - \fx
\partial _{x}\varphi_{\beta} (x,\theta) - \ftheta \partial _{\theta}\varphi
_{\beta}(x,\theta)
\end{equation}
and
\begin{equation} \label{eq:equivalence in Hormander jets}
i\hbar \partial _{\ftheta} a - \partial _{\ftheta} \varphi
_{\beta}(x,\theta; \fx, \ftheta) \eq 0
\end{equation}
Define the action of ${\cal{W}}_{U_{\beta}}$ on the above space by:
\begin{eqnarray}
\fxi :e^{\fih \varphi_{\beta} }a \mapsto e^{\fih \varphi_{\beta}
}(i\hbar \partial _{\fx} a - \partial _{\fx}\varphi _{\beta}a)
\nonumber \\
\fx:e^{\fih \varphi_{\beta} }a \mapsto e^{\fih \varphi_{\beta}
}\fx a
\end{eqnarray}
\end{definition}
\begin{lemma}
The transition functions \eqref{eq:eqivalence for jet module 1}
define a structure of a bundle of ${\cal W}_L \otimes
{\mathbb{K}}$-modules on $J^H_L$. The formula
$$\nabla = (\frac{\partial}{\partial x} - \frac{\fxi}{i\hbar})dx +
(\frac{\partial}{\partial \xi} + \frac{\fx}{i\hbar})d\xi$$ defines
a flat connection on $J^H_L$.
\end{lemma}
{\bf Proof.} Let us check how the above definition differs from
the one given by \eqref{eq:eqivalence for jet module},
\eqref{eq:eqivalence for jet module 1}. There are two differences,
namely the constant term and the linear term of $\varphi (x+\fx,
\theta+\ftheta)$. But these two differences exactly mirror the
differences between the two bundles of algebras with connection, namely $(\pi ^*{\operatorname{jets}}\RDX, \pi ^* \nabla _{\operatorname{can}})$
and $({\cal{W}}_{T^*X}, \nabla)$; they disappear after we modify
the bundle and the connection as in \ref{ss:Differential operators
and the deformation quantization}. Indeed,
\begin{eqnarray}
&\varphi
(x+\fx, \theta+\ftheta) = \varphi (x, \theta) + \fx \varphi _x(x,
\theta) +\ftheta \varphi _{\theta}(x, \theta) + \varphi(x,\theta;
\fx, \ftheta) = \nonumber \\
&=\varphi (x, \theta) + \xi \fx + \varphi(x,\theta; \fx,
\ftheta)\nonumber
\end{eqnarray}
if the point $(x,\theta)$ corresponds to a point of $L$. After
identifying $\pijd$ with $W^{\operatorname{fin}}$ (cf. \eqref{eq:Wfin}), we get a bundle
of $W^{\operatorname{fin}}$-modules with a compatible connection;
after the gauge transformation $\exp (
\frac{1}{i\hbar}\xi \fx)$ is applied to the module, it becomes a
bundle of ${\cal W} \otimes \mathbb{K}$-modules. In coordinates,
the connection is equal to
$$\nabla = \fih d\varphi + (\partial _x - \foh \fxi)dx +
(\partial _{\xi} + \foh \fx)d\xi. $$
But $d\varphi = \alpha $ on
$L$; so, after adding $\foh \xi dx $ to the connection on $W$, the
first term vanishes.
\begin{example} \label{example:jets as W-module}
Let $L$ be given by the equation $\xi ^m = \varphi _{x^m}(x)$ on
$T^*\bbR ^n$. Then sections of the modified jet bundle are formal
expressions
$$e^{\fih (\varphi (x+\fx) - \varphi (x) -\fx \varphi
_x(x))}a(x,\fx)|dx|^{\fot}$$
on which $\fx ^m$ acts by
multiplication and $\xi ^m$ by $i\hbar \partial _{\fx ^m}$. The
connection
$$(\partial _x - \foh \fxi)dx +
(\partial _{\xi} + \foh \fx)d\xi$$
acts at the level of $a(x,\fx)$ by $(\partial _x - \partial
_{\fx})dx$.
\end{example}
\begin{example} \label{example:jets as W-module 2}
Let $L$ be given by the equation $x ^m = -\psi _{\xi^m}(\xi)$ on
$T^*\bbR ^n$. Then sections of the modified jet bundle are formal
expressions
$$e^{\fih (\fx \ftheta + \psi (\xi+\ftheta) - \psi(\xi) -\ftheta
\psi _{\xi}(\xi))}a(\xi,\fx, \ftheta)|d\xi|^{\fot}
\operatorname{mod}\eq$$
where $i\hbar \partial _{\ftheta}(e^{\fih \ldots}a \ldots) \eq
0$.The space of such local sections is isomorphic to the space of
expressions
$$e^{\fih (\fx \ftheta + \psi (\xi+\ftheta) - \psi(\xi) -\ftheta
\psi _{\xi}(\xi))}a(\xi, \ftheta)|d\xi|^{\fot}$$
on which $W$ acts, at the level of the factor $a$, by
$$\fxi ^m: a \mapsto -i\hbar \partial _{\ftheta}a$$
$$\fx ^m: a \mapsto i\hbar \frac{\partial a}{\partial \ftheta} +
\psi (\xi + \ftheta) - \psi (\xi)$$

The connection
$$(\partial _x - \foh \fxi)dx +
(\partial _{\xi} + \foh \fx)d\xi$$
acts at the level of $a(\xi, \ftheta)$ by $(\partial _{\xi} - \partial
_{\ftheta})d\xi$.
\end{example}
The above two examples generalize to the case of any special phase
function \eqref{eq:definition of a phase function}. From this one
deduces
\begin{proposition} \label{prop:hormander jet bundle}
One has for the Lagrangian module $V_L$:
$$V_L \isomoto (J_L^H)^{\nabla} \otimes {\cal E}_{\frac{i}{\hbar}\alpha_L}$$
where the first factor in the right hand side stands for the sheaf of
horizontal sections of $J_L^H$ and $\alpha _L$ is an $\bbR$-valued one-cocycle representing the
cohomology class of $\alpha = \xi dx$ on $L$.
\end{proposition}
Furthermore, from the explicit formulas for the transition
functions one observes the following
\begin{proposition}
\label{prop:filtration on Hormander jets} The fiber of $J_L^H$ is
isomorphic to $\bbC[[\hbar, \fx]] \otimes _{\bbC [[\hbar]]} \bbK$.
If we put $|\fx| = 1$ and $|\hbar | = 2$ and consider the
filtration $F^m = \prod _{p \geq m}\{a|\,|a| = p\}$, then this
filtration induces a filtration on $J_L^H$ which is compatible
with the similar filtration on $W$.
\end{proposition}
\section{The main statement} \label{section:main}
\subsection{} We have defined a deformation quantization
${\mathbb{A}}_{T^*X}$ (subsection \ref{ss:Differential operators and the deformation
quantization}) and a sheaf of $\mathbb{A}_{T^*X} \otimes
\mathbb{K}$-modules $V_L$ (section \ref{s:Hormander in deformation quantization}). Now, using Darboux-Weinstein theorem,
we can identify a neighborhood of $L$ in $T^*X$ with a
neighborhood ${\stackrel{0}{T^*}}L$ of $L$ in $T^*L$. We can construct the algebra
${\mathbb{A}}^0 _{T^*L}$ and the module $V^0_L$ using this
identification, and choosing the zero section $L$ as a Lagrangian
submanifold of $T^*L$.
\begin{prop}\label{prop:can up to in}
There exists an isomorphism of algebras on ${\stackrel{0}{T^*}}L$
$$ {\mathbb{A}}_{T^*X} \isomoto  {\mathbb{A}}^0_{T^*L}$$
which is canonical up to a canonical inner automorphism.
\end{prop}
Indeed, because of the remark in the end of the proof of \ref{prop:theta on T*X}, one can apply Theorem \ref{thm:modified Fedosov construction}.
\begin{thm} \label{thm:main}
Let us identify the algebras ${\mathbb{A}}_{T^*X}$ and ${\mathbb{A}}^0_{T^*L}$ using Proposition \ref{prop:can up to in}. There exists an isomorphism of modules
$$ V_L \isomoto V^0_L \otimes _{\bbC[[\hbar]]} {\cal E}_{\fih \alpha _L +
\frac{\pi i}{2}\mu _L}$$
where $\alpha _L$ is an $\bbR$-valued one-cocycle representing the
cohomology class of $\alpha = \xi dx$ on $L$, $\mu _L$ is a
$\bbZ$-valued one-cocycle representing the Maslov class of $L$,
and ${\cal E}_{\fih \alpha _L + \frac{\pi i}{2}\mu _L}$ is the
$\bbK$-valued local system on $L$ with the transition functions
$\exp({\fih \alpha _L + \frac{\pi i}{2}\mu _L})$.
\end{thm}
{\bf{Proof.}} We have proven that ${\mathbb{A}}_{T^*X}$ is
isomorphic to the algebra of horizontal sections of the bundle of
algebras ${\cal{W}}_{T^*X}$, and there is a compatible
isomorphism of $V_L$ to the module of horizontal sections of the
bundle of modules $J_L^H$. Similarly, ${\mathbb{A}}^0_{T^*L}$ is
isomorphic to the algebra of horizontal sections of the bundle of
algebras ${\cal{W}}^0_{T^*L}$, and that there is a compatible
isomorphism of $V^0_L$ to the module of horizontal sections of the
bundle of modules $J_L^{H, 0}$. (The algebra isomorphisms are canonical up to a canonical inner automorphism). Therefore the statement of the
theorem follows from
\begin{proposition} \label{prop:main}
1). There is a connection-preserving isomorphism of bundles of
algebras on ${\stackrel{0}{T^*}}L$
$${\cal{W}}_{T^*X} \isomoto {\cal{W}}^0_{T^*L}$$
which is canonical up to a conjugation by a canonical invertible horizontal element.

2).There is a connection-preserving isomorphism of bundles of
modules
$$J_L^H \isomoto J_L^{H,0} \otimes _{\bbC[[\hbar]]} {\cal E}_{
\frac{\pi i}{2}\mu _L}$$
compatible with the above isomorphism of bundles of algebras.
\end{proposition}
{\bf{Proof of Proposition.}} All our local phase functions will be of the special form \eqref{eq:definition of a phase function}. Start with the transition functions for the bundle of algebras ${\cal{W}}_{T^*X}$; 
$$\tG _{\alpha \beta} : U_{\alpha} \cap U_{\beta} \to K \to \tG _{\geq 0}$$
Similarly, consider the lifted transition functions $\tG ^0 _{\alpha \beta}$ for the bundle ${\cal{W}}^0 _{T^*L}$. The group $K$ and its embedding to $\tG _{\geq 0}$ are explained in the end of \ref{ss:Differential operators and the deformation
quantization}. The lifted Fedosov connection, in local coordinates, is given by the formula
\begin{equation} \label{eq: the canonical connection} 
(\partial / \partial x - \fxi /i\hbar) dx + (\partial / \partial \xi + \fx /i\hbar)d\xi
\end{equation}
Here and below we call the connection given by this formula {\em the canonical connection.}

Now replace $\tG _{\alpha \beta}$ by an equivalent set of transition functions
$$\tG ^{\operatorname{new}} _{\alpha \beta} = \tH _{\alpha} \tG _{\alpha \beta} \tH _{\beta}^{-1}$$ where 
$$\tH_{\alpha}: U_{\alpha} \to \tG _{\geq 0}$$
are defined as follows.

Let $L\cap {U_{\alpha}}$ be given by \eqref{eq:definition of a phase function}. Define

$$\sigma_{\alpha}=\exp\fih( F(x_1 + \fx _1, \xi _2 + \fxi _2)- F(x_1, \xi _2 )-F_{x_1}(x_1 , \xi _2 )\fx _1 -F_{\xi_2}(x_1 , \xi _2 )\fxi _2)$$
in $\tG _{\geq 0}$.
Now consider the local coordinate change 
$$(x_1, x_2) \mapsto (x_1, \xi_2); \;\; (\xi _1, \xi _2) \mapsto (\xi _1, -x_2)$$
and a 
symplectic transformation (the partial Fourier transform)
$${\bf F}_{\alpha}: \fx \mapsto (\fx_1, \fxi_2); \;\; \fxi \mapsto (\fxi _1, -\fx_2)$$
We fix liftings ${\widetilde {\bf F}}_{\alpha}$ to $\TSP$ (counterclockwise rotation in $(\fx _2 , \fxi _2)$ space). Put 
$$\tH _{\alpha} = {\widetilde {\bf F}_{\alpha}}\sigma_{\alpha}$$

Note that the above formula is precisely the Maslov canonical operator, defined here at the jet level.

We get the new bundle of algebras, which we denote by
${\cal{W}}^{\operatorname{new}}_{T^*X}$. The connection
on $J^{H}_L$ is given by the same formula as the canonical
connection. The action of ${\cal{W}}^{\operatorname{new}}_{T^*X}$
on $J^{H}_L$ is, in our new local coordinates, the standard one:
$\fx$ acts by multiplication,and $\fxi$ by $i\hbar
\frac{\partial}{\partial{\fx}}$. Note that, because of this, the transition functions of the module $J^L_H$ determine the transition functions of the bundle of algebras ${\cal{W}}^{\operatorname{new}}_{T^*X}$. The same is true about $J^{H,0}_L$ and ${\cal{W}}^{0}_{T^*L}$.
 
We claim that:

1) The transition functions $\tG^{\operatorname{new}}_ {\beta \gamma}$ take values in the subgroup $P$ (cf. Lemma \ref{lemma:estabilizadores}; note also that $K$ is a subgroup of $P$).

2) The image of $\tG^{\operatorname{new}}_ {\beta \gamma}$ in $P/N$ is equal to the image of $\mu _{\beta \gamma}\tG^{0}_ {\beta \gamma}$.

Here $\mu _{\beta \gamma}$ is the cocycle representing the Maslov class, as in \eqref{eq:Maslov cocycle 2}, with values in $\bbZ \subset \tG _{\geq 0}$.

3) The transition functions of the bundle of modules $J^H_L$ are equal to $\exp({\bf {\alpha}}_{\beta \gamma}) w(\tG^{0}_ {\beta \gamma})$ where ${\bf \alpha}_{\beta \gamma}$ is the specific cocycle representing the 1-cohomology class $\alpha$ of $L$ as in \eqref{eq:alpha}
$$w: P \to \operatorname{Aut} (\bbC[[\fx, \hbar]])$$
is the restriction of the degenerate Weil representation
$$w: \tG_{\geq 0} \to \operatorname{Aut}({\widehat{V}}^0_{\operatorname{Weil}})$$
to the subgroup preserving the subspace $V_{T=0} = \bbC[[\fx, \hbar]]$.

To prove the claim, observe first the following.

A) The transition functions $\tG^{\operatorname{new}}_ {\beta \gamma}$ and $\tG^{0}_ {\beta \gamma}$ are the same if they correspond to a coordinate change $g_{\alpha(\beta) \alpha( \gamma)}$ on $X$, and $U_{\beta}$, $U_{\gamma}$ are such that the projections of $U_{\beta} \cap L$ and $U_{\gamma} \cap L$ to the base are
bijective. Similarly, the transition functions of $J^{H}_L$ and $J^{H,0}_L$ are the same, and are the image of the above under $w$.

B) The same is true modulo $N$ if the transition functions $G^{\operatorname{new}}_{\beta \gamma}$ correspond to a change of subdivision $x = (x_1, x_2)$. This follows from the formula \eqref{eq:statfaza}, or rather from its version for the power series (cf. \cite{K}). More precisely, the transition functions $\tG^{\operatorname{new}}_ {\beta \gamma}$ are given by
$$\mu_ {\beta \gamma}\exp (\frac{1}{i\hbar}\sum \hbar ^{h_1 (\Gamma)}  c_{\Gamma})$$
where $\Gamma$ are all connected graphs; the sum of the terms with ${h_1 (\Gamma)}=0,\;1$ is  exactly the transition functions $\tG^{0}_ {\beta \gamma}$. Similarly, the transition functions of the module $J_L^H$ are given by 
$$\exp(\frac{\pi i}{2}\mu_ {\beta \gamma}) (\frac{1}{i\hbar}\sum \hbar ^{h_1 (\Gamma)}  c_{\Gamma})$$
and the transition functions of the module $J^{H,0}_L$ are given by the sum of the terms with ${h_1 (\Gamma)}=0,\;1$. Again, we see directly that the transition functions of the bundle of modules are the image under $w$ of the transition function of the bundle of algebras.

C) It remains to compare our transition functions for the rest of coordinate changes, namely, when the coordinate change corresponds to a coordinate change on the base and the subdivision $x = (x_1, x_2)$ has $n_2 > 0$. Observe that all the transition functions that we are considering are given by universal formulas in terms of two jets of coordinate systems on the base, a jet of a Lagrangian, and two subdivisions $x = (x_1, x_2)$. On an open dense subset, the transition functions of the type C) can be expressed through the transition functions of types A), B). But, because of the above arguments, all the equalities that we are proving are true on an open dense subset; therefore, they are true everywhere.

We see now that the transition functions $\tG^{\operatorname{new}}_ {\beta \gamma}$ and $\tG^{0}_ {\beta \gamma}$ differ by a \v{C}ech one-cocycle with values in $N$, invariant under the canonical connection $(\partial / \partial x - \fxi /i\hbar) dx + (\partial / \partial \xi + \fx /i\hbar)d\xi$. But it is easy to see that any such cocycle is cohomologous to the identity. In fact, finding a zero-cochain of which it is a coboundary reduces to an iterative procedure whose individual steps are to trivialize a \v{C}ech one-cocycle with coefficients in a sheaf of smooth sections of a $C^{\infty}$ vector bundle.

\subsection{The main statement in the Fedosov form} We finish by
identifying the Maslov-H\"{o}rmander construction in deformation
quantization in Fedosov terms. Our first goal is to determine the
structure of the associated graded module $\operatorname{gr}_F
J_L^H$.
\subsubsection{The flat bundle $(W/WT_L ^{\perp})\otimes |\Omega
_L|^{\fot}$} For any symplectic $M$ and any
Lagrangian submanifold $L$, let $T_L^{\perp}$ be the conormal
bundle of $L$, viewed as a subbundle of the Weyl bundle $W_L$.
Obviously, $W/WT_L ^{\perp}$ is a $W$-module. Let $\nabla$ be a Fedosov connection compatible with $L$. Choose a flat connection on the bundle of half-densities$ |\Omega
_L|^{\fot}$. 
The tensor product $(W/WT_L ^{\perp})\otimes |\Omega _L|^{\fot}$ becomes a $W$-module with a flat connection which is compatible with the Fedosov
connection on $W$.

The following is easy to see from the explicit definition of
$J_L^H$.
\begin{proposition} \label{prop:grJ}
There is an isomorphism of bundles of $W \otimes \bbK$-modules
$$\operatorname{gr}_F (J_L^H \otimes _{\bbC[[\hbar]]} {\cal E}^{-1}_{\fih \alpha _L +
\frac{\pi i}{2}\mu _L})\isomoto (W/WT_L ^{\perp}\otimes |\Omega
_L|^{\fot}) $$
where $\alpha _L$ is an $\bbR$-valued one-cocycle representing the
cohomology class of $\alpha = \xi dx$ on $L$, $\mu _L$ is a
$\bbZ$-valued one-cocycle representing the Maslov class of $L$,
and ${\cal E}_{\fih \alpha _L + \frac{\pi i}{2}\mu _L}$ is the
$\bbK$-valued local system on $L$ with the transition functions
$\exp({\fih \alpha _L + \frac{\pi i}{2}\mu _L})$.
\end{proposition}

\begin{thm} \label{thm:main1}1) The deformed algebra $\bbA _{T^*X}$ is isomorphic to the algebra of horizontal sections of the bundle $W$. This isomorphism is canonical up to a canonical inner automorphism. 

2) Under the identification from the statement 1), the Lagrangian module $V_L^H$ is isomorphic to the sheaf of
horizontal sections of $(W/WT_L ^{\perp}\otimes |\Omega
_L|^{\fot}) \otimes _{\bbC[[\hbar]]} {\cal E}_{\fih \alpha _L +
\frac{\pi i}{2}\mu _L}$.
\end{thm}
Statement 1) follows from Theorem \ref{thm:modified Fedosov construction}. Statement 2) follows from Proposition \ref{prop:main} and the following
\begin{proposition} \label{prop:main1} The bundles of $W_L$-modules
with connections are isomorphic:
$$J_L^H \isomoto (W/WT_L ^{\perp}\otimes |\Omega
_L|^{\fot}) \otimes _{\bbC[[\hbar]]} {\cal E}_{\fih \alpha _L +
\frac{\pi i}{2}\mu _L}$$
\end{proposition}
{\bf Proof.} Proposition \ref{prop:main} reduces this statement to the case when $L$ is the zero section of the cotangent bundle, where it is easy to see explicitly.


\begin{thebibliography}{ABCDEF}

\bibitem[BFFLS]{BFFLS} F.~Bayen, M.~Flato, C.~Fronsdal, A.~Lichnerowicz,
D.~Sternheimer, {\em Deformation theory and quantization}, Ann. Phys. {\bf
111} (1977). p. 61-151

\bibitem[BK]{BK} R.~Bezrukavnikov, D.~Kaledin, {\em Fedosov quantization in algebraic context}, math.AG/0309290.

\bibitem[Bo]{Bo} M.~Bordemann, {\em (Bi)modules, morphismes et r\'{e}duction des star-produits: le cas symplectique, feuilletages et obstructions}, math.QA/0403334v1 (2004).

\bibitem[BNT]{BNT} P.~Bressler, R.~ Nest, B.~ Tsygan, {\em Riemann-Roch
theorems via deformation quantization. I, II},  Adv. Math.  {\bf 167}
(2002),  no. 1, 1--25, 26--73.

\bibitem[BS]{BS} P.~Bressler, Y.~Soibelman, {\em Mirror symmetry and
deformation quantization}, hep-th 0202128
\bibitem[DP]{DP} A.~D'Agnolo, P.~Polescello, {\em Stacks of twisted modules and integral transforms}, math.AG/0307387
\bibitem[Fe]{Fe} B.~Fedosov, {\em Deformation Quantization and Index
Theorem}, Akademie Verlag, 1994.

\bibitem[GS]{GS} V.~Guillemin, S.~Sternberg,  {\em Symplectic techniques in
physics}, second edition, Cambridge University Press, Cambridge, 1990.

\bibitem[GS1]{GS1} V.~Guillemin, S.~Sternberg, {\em Some problems in integral geometry and some related problems in microlocal analysis}, Amer. J. Math. {\bf 101} (1979), 915-955.

\bibitem[GU]{GU} V.~Guillemin, A.~Uribe, {\em Reduction and the trace formula}, Journal Differential Geometry {\bf 32} (1990), 315-347.
\bibitem[H]{H} L.~H\"{o}rmander, {\em Fourier integral operators I}, Acta
Mathematica {\bf 127}  (1971), no. 1-2, 79--183.

\bibitem[K]{K} D.~Kazhdan, in: Quantum fields and strings: a course for
mathematicians, AMS, vol. 1, 1996.
\bibitem[Kas]{Kas} M.~Kashiwara,{\em Quantization of contact manifolds}, Publ. Res. Inst. Math. Sci. {\bf 32}, 1 (1996), 1-7 
\bibitem[Kar]{Kar} M.~Karasev, 
{\em Quantization and coherent states over Lagrangian submanifolds},
Russian J. Math. Phys. {\bf 3} (1995), no. 3, 393--400.
\bibitem[KS]{KS} M.~Kontsevich,Y.~Soibelman,{\em Homological mirror symmetry
and torus fibrations},  Symplectic geometry and mirror symmetry (Seoul,
2000),  203--263, World Sci. Publishing, River Edge, NJ, 2001.

\bibitem[L]{L} J.~Leray, {\em Lagrangian analysis and quantum mechanics,}
Studies in applied mathematics, {\bf 7--9}, Adv. Math. Suppl. Stud., 8,
Academic Press, New York, 1983.

\bibitem[M]{M} V.~Maslov, {\em Operational methods}, Mir, Moscow, 1976.

\bibitem[NT2]{NT} R.~Nest, B.~Tsygan, {\em Formal versus analytic index
theorems}, IMRN {\bf 11}, 1996, 557-564

\bibitem[MSS]{MSS} A.~Mishchenko, V.~Shatalov, B.~Sternin,{\em Lagrangian manifolds and Maslov operator}, Springer Lectures in Soviet Mathematics, Springer, 1990.
\bibitem[NSS]{NSS} V.~Nazaikinskii, B.-W.~Schulze, B.~Sternin, {\em Quantization methods in differential equations}, Differential and Integral Equations and their applications, Taylor and Francis, Ltd, London, 2002.

\bibitem[NT1]{NT1} R.~Nest, B.~Tsygan, {\em Deformations of symplectic Lie
algebroids, deformations of holomorphic symplectic structures, and index
theorems}, Asian J. Math. {\bf 5}  (2001),  no. 4, 599--635

\bibitem[PS]{PS} P.~Polescello, P.~Schapira, {\em Stack of deformation quantization modules on complex symplectic manifolds}, math.AG/0305171.

\bibitem[Se]{Se} P.Seidel, {\em Graded Lagrangian submanifolds}, Bull.
Soc.Math.France,{\bf 128}  (2000),  no. 1, 103--149.

\bibitem[W]{W} S.~Waldmann, {\em On the representation theory of deformation quantization}, in: G.~Halbout (ed.), Deformation Quantization, tome 1 in: IRMA Letters in Mathematics and Mathematical Physics, De Gruyter, Berlin - New York (2002), 107--133.


\end{thebibliography}
\end{document}